\documentclass[a4paper,11pt]{article}
\pdfoutput=1 

\usepackage{jcappub} 

\usepackage[T1]{fontenc} 

\DeclareMathOperator\arctanh{arctanh}
\usepackage{float}

\title{\boldmath Addressing the Hubble and $S_8$ Tensions with a Kinetically Mixed Dark Sector}

\author[1]{Stephon Alexander}

\author[2]{Heliudson Bernardo}

\author[1]{Michael W. Toomey}

\affiliation[1]{Brown Theoretical Physics Center and Department of Physics, Brown University,\protect\\  Providence, RI 02912, USA}

\affiliation[2]{Department of Physics, McGill University,\protect\\ Montreal, QC, H3A 2T8, Canada}

\emailAdd{stephon\_alexander@brown.edu}
\emailAdd{heliudson@hep.physics.mcgill.ca}
\emailAdd{michael\_toomey@brown.edu}

\abstract{We present a kinetically mixed dark sector (KMIX) model to address the Hubble and $S_8$ tensions. Inspired from string theory, our model includes two fields: an axion, which plays a role similar to the scalar field in early dark energy models, and a dilaton. This theory differs from other axio-dilaton models aimed at the Hubble tension in that there is necessarily kinetic mixing between the two fields which allows for efficient energy transfer from the axion into the dilaton which has $w\approx1$. As a direct consequence of these dynamics, we find the model does not need to resort to a fine-tuned potential to solve the Hubble tension and naturally accommodates a standard axion potential. Furthermore, the axion will necessarily makeup a small (fuzzy) fraction of $\Omega_{\rm cdm}$ once it begins to oscillate at the bottom of its potential and will suppress the growth of perturbations on scales sensitive to $S_8$. Interestingly, the scale of the potential for the dilaton has to be small, $\lesssim \mathcal{O}(10~{\rm meV})^4$, suggesting the possibility for a connection to dark energy. Implementing the dynamics for the background and perturbations in a modified Boltzmann code we calculate the CMB and matter power spectra for our theory. Exploring the parameter space of our model, we find regions which can accommodate a $\sim 10\%$ increase in $H_0$ from the Planck inferred value and $S_8$ values that are consistent with large-scale structure constraints. }

\begin{document}
\maketitle
\flushbottom

\section{Introduction and Motivation}
\label{sec:intro}

The tension between local and global measurements of $H_0$ -- the first ever measured parameter of the $\Lambda\text{CDM}$ model -- has arrived at the $4-6\sigma$ range, depending on the combined data sets \cite{Verde:2019ivm, Planck:2018vyg, Riess:2019cxk, Abdalla:2022yfr, DiValentino:2021izs}. In fact, the $H_0$ value inferred from the Planck 2018 cosmic microwave background (CMB) data disagrees with local measurements by the SH$0$ES collaboration at a statistical significance of 5$\sigma$ \cite{Riess:2021jrx}. While the SH$0$ES result relies on cepheid calibration of the cosmic distance ladder \cite{Sandage:2006cv}, its is not clear that changing the calibration method alleviates the tension \cite{Freedman:2020dne,Huang:2019yhh} due to a loss in precision.

If not due to non-obvious systematic effects in the measurements, this ``Hubble tension'' requires modifying the standard cosmological model, a challenging task for the precision fits of other parameters and the cosmological concordance between data across different scales \cite{Ivanov:2020ril}. In particular, the angular scale $\theta_s$ of the comoving sound horizon $r_s$ at decoupling is measured from the acoustic peaks of the CMB power spectrum at a precision of $0.03\%$ \cite{Planck:2018vyg} and its value is fixed by the $\Lambda$CDM parameters through its dependence on the sound horizon and the angular diameter distance, 
\begin{equation}
    \theta_s = \frac{r_s(z_*)}{D_A(z_*)}, \quad r_s(z_*) = \int_{z_*}^{\infty} \frac{c_s}{H(z)}dz\;, \quad D_A(z_*) = \int_0^{z_*}\frac{1}{H(z)}dz,
\end{equation}
where $z_*$ is the redshift at photon-baryon decoupling. Hence, shifts in the CMB-inferred $H_0$ towards local-measured values should be followed by a change in $r_s$ so as to keep $\theta_s$ unchanged. Comparing Planck 2018 and SH$0$ES central values for $H_0$, we need an increase of $\sim 10\%$ to solve the Hubble tension. This entails a decrease in $r_s(z_*)$ by the same factor. But since the integral defining $r_s(z_*)$ is dominated by the cosmological evolution just prior to recombination, the most efficient way to shrink it is by increasing the background energy density around $z_*$. 

There are currently several proposals for increasing $H(z_*)$ to solve the Hubble tension (see \cite{DiValentino:2021izs,Schoneberg:2021qvd} and references therein). ``Early Dark Energy'' (EDE) models are a subclass of those that postulates the existence of a subdominant energy density component that peaks before $z_*$ but that quickly redshifts shortly afterwards \cite{Karwal:2016vyq,Poulin:2018cxd,Agrawal:2019lmo,Alexander:2019rsc,Lin:2019qug,Perez:2020cwa,Niedermann:2019olb,Sakstein:2019fmf, Ye:2020btb,Freese:2021rjq,Braglia:2020bym,Gogoi:2020qif}. This idea is often implemented after introducing a canonically normalized scalar field $\phi$ with a potential $V(\phi)$ that has a curvature scale $\sqrt{V_{\phi\phi}}$ of the order of $H(z_*)$ such that it is held constant by Hubble friction at early times. Around before $z_*$, typically closer to matter-radiation equality, the field rolls down its potential and the subsequent late time evolution depends on $V(\phi)$. For instance, coherent (an)harmonic oscillations around $\sim \phi^{2n}$ minima makes the energy density in the field redshift as $a^{-\frac{6n}{n+1}}$ \cite{Turner:1983he}, so if $n > 2$ the energy density in the field redshifts faster than radiation. Moreover, since $H(z_*)\sim 10^{-28} $ eV, an ultra-light field is needed.

From a theoretical perspective, the EDE scalar field should be an ultralight axion whose potential is generated by non-perturbative effects, $V \sim m_\phi^2 f^2_\phi(1 - \cos \phi/f_\phi)$, where $m_\phi$ and $f_\phi$ are the axion mass and decay constant, respectively. However, if there is no energy transfer from $\phi$ to another component that redshifts faster than radiation, higher harmonics need to dominate the instanton expansion to recover the $\Lambda$CDM cosmological evolution after recombination. 
Potentials of this form, $V\sim m_\phi^2 f^2_\phi (1- \cos( \phi/f_\phi))^n$, were studied in \cite{Poulin:2018dzj} and constitute 3 extra parameters relative to $\Lambda$CDM for a choice of $n$: the initial displacement of the field $\theta_i :=\phi_i/f_\phi$, the redshift $z_c$ at which the energy density fraction in $\phi$ is maximum, and its value $f_{\text{EDE}}(z_c)$ at the maximum.  It was shown in \cite{Smith:2019ihp} that data has a preference for $n=3$ models. However, having a controlled instanton expansion dominated by a higher harmonic term requires fine-tuning the expansion coefficients. These are fixed by the details of the non-perturbative physics giving rise to the expansion. However, this a theoretical challenge for EDE models with $n\geq2$ and currently there are no concrete derivations for such a potential. Phenomenologically, this is a mild issue as we should let data constrain $V(\phi)$. 

Combining CMB and large-scale structure (LSS) data yields stringent constraints on EDE models. The $\Lambda$CDM parameter shift necessary to accommodate the EDE phase exacerbates the tension between the amplitude of density fluctuations at late times as inferred from CMB and LSS data \cite{Hill:2020osr, Ivanov:2020ril, DAmico:2020ods, Herold:2021ksg,Smith:2020rxx,Hill:2021yec}. Although for a short period, the EDE component slightly reduces the growth of perturbations such that the $\Lambda$CDM parameters are shifted to compensate for that. In particular, the physical cold dark matter density increases leading to a slightly larger value of $S_8$ as compared with $\Lambda$CDM \cite{Hill:2020osr}. Hence, generically, any model that works on these grounds will increase the $S_8$ tension between early and late-time data sets \cite{Vagnozzi:2021gjh}. This tension is even present in $\Lambda$CDM with the Planck inferred value at $S_8 = 0.834^{+0.016}_{-0.016}$ \cite{Planck:2018vyg} and current constraints from large-scale structure at $S_8 = 0.800^{+0.029}_{-0.028}$ for HSC Y1  \cite{HSC:2018mrq}, $S_8 = 0.759^{+0.024}_{-0.021}$ for KiDS-1000 \cite{KiDS:2020suj}, and $S_8 = 0.772^{+0.018}_{-0.017}$ for DES Y3 \cite{DES:2021vln} 
- this is the ``$S_8$ tension.'' \footnote{It should be pointed out that this tension is less clear and could be a statistical fluctuation \cite{Nunes:2021ipq}.}

The challenges described above motivate searching for alternatives to single-field EDE models. Indeed, many phenomenological extensions of the standard EDE dynamics have been studied in the literature. These include a portion of dark matter decaying at late times \cite{Clark:2021hlo}, the addition of a second, light axion contribution to dark matter \cite{Allali:2021azp,Ye:2021iwa}, massive neutrinos \cite{Reeves:2022aoi}, and a coupling of dark matter to the EDE scalar motivated by the swampland distance conjecture \cite{McDonough:2021pdg}, all with the aim to suppress structure formation. In a similar spirit, proposals for dissipating the EDE density into other components, which naturally redshift faster than radiation, have also been studied \cite{Alexander:2019rsc,Lin:2019qug,Niedermann:2019olb,Berghaus:2019cls,Berghaus:2022cwf}. In particular, in \cite{Berghaus:2019cls} a dissipative axion model was proposed in which the energy gets transferred to dark radiation, the latter acting as an extra friction in the $\phi$ equation of motion. Thermal friction was also shown to alleviate the Hubble and LSS tensions \cite{Berghaus:2022cwf}, but the data fit is such that the energy transfer happens at redshifts much higher than matter-radiation equality, 
deep into radiation domination where data has lost sensitivity to the transition and the model asymptotes to an extra radiation solution.
One can of course also consider abandoning EDE-based models altogether. Indeed, models inspired by supersymmetry which modify $N_{\rm eff}$ have seen success in not only addressing the Hubble tension \cite{Aloni:2021eaq} but also the $S_8$ tension \cite{Joseph:2022jsf}. While the latter model is a step in the right direction, there is a clear need for more well-motivated models that have the capacity to naturally address both tensions.

Motivated by string theory effective actions, we propose a two-field model that shares some similarities with usual EDE and thermal friction models: on one hand, there is still a phase of ``early dark energy'' which modifies the size of the sound horizon and on the other hand there is energy transfer from one component to another, which also contributes as friction to the evolution of the energy injecting field. However, no thermal effects are needed, the energy transfer is due to a necessary kinetic coupling between the fields, something that to the best of our knowledge has not been considered in the literature. Such a kinetic coupling is a generic feature of the effective theory for complex moduli fields in string compactification. More specifically, our starting action is
\begin{equation}\label{scalarsectoraction}
    S \supset \int d^4x \sqrt{-g}\left[- \frac{1}{2}g^{\mu\nu}\partial_\mu \chi \partial_\nu \chi - \frac{1}{2}f(\chi)g^{\mu\nu}\partial_\mu \phi \partial_\nu \phi - V(\chi, \phi)\right],
\end{equation}
where the kinetic mixing function $f(\chi)$ is assumed to be exponential -- the natural case when $\chi$ is the modulus associated to extra dimensions -- and the potential $V(\chi, \phi)$ does not need to be fine tuned (see Appendix \ref{string-append} for string theoretic derivation). As we shall discuss in the next section, the kinetic coupling between $\phi$ and $\chi$ produces a source term in the equation of motion for $\chi$ and a friction term in the equation of motion for $\phi$, making the energy transfer possible but damped by friction. Furthermore, an unavoidable consequence of our model is a small (fuzzy) axion contribution to dark matter which naturally suppresses structure on scales sensitive to $S_8$.

Promisingly, the string axio-dilaton modulus has the kinetic structure in Eq.~\ref{scalarsectoraction} regardless of the model considered. In fact, in \cite{Alexander:2019rsc}, a string theory inspired axio-dilaton model was shown to resolve the Hubble tension. A major difference being that the dilaton, and not the axion, played the role of dark energy. The dilaton is instead stabilized by the axion until the mass of the latter falls below $H$. At that point the dilaton becomes destabilized and enters fast-roll, thus diluting its energy density as $a^{-6}$. However, since the kinetic coupling was not considered in \cite{Alexander:2019rsc}, the model relied on the choice of an interaction potential to dilute the dilaton's energy density. This is in contrast to our proposed model,\footnote{Note that although we call $\chi$ a ``dilaton'', it is not necessarily the string dilaton which sets the string coupling. In string theory, a pair like $(\phi,\chi)$ might also correspond to the size of cycles in the compact internal manifold and fluxes on those cycles, respectively.} which works for natural potentials such as mass terms or the first harmonic of an instanton expansion.

In Section~\ref{background} we discuss the background dynamics of the scalar field sector where the action of the form in Eq.~\ref{scalarsectoraction} is studied. We show that, in a homogeneous and isotropic background, the energy transfer from $\phi$ to $\chi$ is generic and that the $\chi$-component redshifts as stiff matter as long as its potential is negligible. Using a modified Boltzmann code we then study the background dynamics for our theory numerically. In Section \ref{pert} we derive the linearized, perturbed Klein-Gordon equations for our theory and compare with those for EDE. We show that $\phi$ will inevitably contribute a small fraction to the dark matter density, naturally suppressing the growth of structure on scales sensitive to $S_8$. Using our modified Boltzmann code we then calculate the CMB and matter power spectra to show that there are regions of parameter space in our model which have the right dynamics to resolve the Hubble and $S_8$ tensions. In Section~\ref{DnC} we discuss the implications of our theory and directions for future work.

\paragraph{Notation and conventions:}If not stated otherwise, we use Planckian units $M_{\text{Pl}} = 8\pi G \equiv 1$ and work in the mostly plus metric signature. We denote $f_\phi$ the axion decay constant and $f_\text{KMIX}(z)$ the fraction of energy density in the dark sector fields at the redshift $z$.

\section{Background dynamics}\label{background}

Consider the scalar-sector action
\begin{equation}
    S[\chi,\phi] = -\int d^4x \sqrt{-g}\left[\frac{1}{2}g^{\mu\nu}\partial_\mu \chi \partial_\nu \chi + \frac{1}{2}f(\chi)g^{\mu\nu}\partial_\mu \phi \partial_\nu \phi + V(\chi, \phi)\right],
    \label{action}
\end{equation}
with $f(\chi)> 0$. In an arbitrary background, its equations of motion are
\begin{subequations}
\begin{align}
    \nabla^2 \chi - \frac{1}{2}f_{\chi}(\chi)g^{\mu\nu}\partial_\mu \phi \partial_\nu \phi - V_\chi = 0, \\
    \nabla_\mu\left[f(\chi) g^{\mu\nu}\partial_\nu \phi\right] - V_\phi = 0,
\end{align}
\end{subequations}
and for homogeneous fields in a flat FLRW metric in comoving coordinates, we have
\begin{subequations}
\begin{align}
    \Ddot{\phi} + \left(3H + \frac{f_\chi}{f} \Dot{\chi}\right)\Dot{\phi} + \frac{1}{f} V_\phi &= 0,\label{phi_eom}\\
    \Ddot{\chi} + 3 H \Dot{\chi} - \frac{1}{2}f_\chi \Dot{\phi}^2 + V_\chi &= 0,
    \label{chi_eom}
\end{align}
\end{subequations}
where we denote partial derivatives with respect to the fields by a subscript. We see that for a non-trivial kinetic coupling function, $f(\chi)$, there is a friction term for $\phi$ that is proportional to $\dot{\chi}$ and a source term for $\chi$ which is proportional to $\dot{\phi}^2$. So, if $\dot{\phi}\neq 0 $, the energy in the $\phi$-component of the system gets transferred to the $\chi$-component. 

\subsection{Analytics}
To understand the dynamics of the fields with the kinetic coupling, let us consider the system in flat space and $V = 0$, but still assume homogeneous fields. With those assumptions, the system is integrable since there are two conserved quantities associated to translations in time and in $\phi$,
\begin{subequations}\label{integrablelimit}
\begin{align}
    f(\chi)\dot{\phi} &= p,\\
    f(\chi)\frac{\dot{\phi}^2}{2} + \frac{1}{2}\dot{\chi}^2 &= E,
\end{align}
\end{subequations}
where $p$ and $E$ are constants. These equations imply that
\begin{equation}
    \frac{1}{2}\dot{\chi}^2 + \frac{p^2}{2} \frac{1}{f(\chi)} = E,
\end{equation}
and so the dynamics for $\chi(t)$ resembles that of an one-dimensional system with an effective potential $p^2f^{-1}/2$. If $\phi$ has no initial velocity, $p=0$ and the solution for $\chi$ is a linear function of $t$, while $\phi$ is constant. However, if $\phi$ has a non-trivial initial kinetic energy then $p\neq 0$ and the motion for $\chi$ is determined from
\begin{equation}
    \dot{\chi} = \pm \sqrt{2E - p^2f(\chi)^{-1}}.
\end{equation}
So, if $f_\chi >0$ ($f_\chi<0$) for all $\chi$, then the physical $\chi(t)$ motion is restricted to values $\chi(t)\geq \chi_c$ ($\chi(t)\leq \chi_c$), where $\chi_c$ satisfies $2Ef(\chi_c) = p^2$. In both cases, $|\dot{\chi}| \to \sqrt{2E}$ and $\dot{\phi} \to 0$ asymptotically, with a small period of energy transfer from $\chi$ to $\phi$ if $f_\chi <0$. If $f(\chi)$ has a minimum, the motion can be periodic, with a cyclic energy exchange between $\phi$ and $\chi$.

For the case we are interested in, $f(\chi) = e^{\lambda \chi}$, the analysis above show us that the energy in $\phi$ gets transferred to $\chi$. In fact, assuming this functional form for $f(\chi)$ and $p\neq 0$, we can integrate the equations of motion exactly,
\begin{subequations}
\begin{align}
    \phi(t) &= \phi_0 +\frac{2\sqrt{2E}}{\lambda p}\left\{\tanh\left[\arctanh A + \sqrt{\frac{E}{2}}\lambda(t-t_0) \right]-A\right\},\\
    \chi(t) &= -\frac{1}{\lambda}\ln \left\{\frac{2E}{p^2}\left[1 - \tanh^2 \left[\arctanh{A}+ \sqrt{\frac{E}{2}}\lambda (t-t_0)\right]\right]\right\},
\end{align}
\end{subequations}
where $A = \sqrt{1-\frac{p^2}{2E}e^{-\lambda \chi_0}}$ and we denoted $\phi_0$ and $\chi_0$ the value of the fields at $t_0$. From this result, one can check that $\dot{\phi}\to 0$ and $|\dot{\chi}|\to \sqrt{2E}$ as $|\chi(t)|$ increases.

From the discussion so far, we conclude that the energy transfer is a generic feature of kinetic coupling. For the rest of this section, we study
whether this can dissipate the early dark energy density and provide a solution to the Hubble tension. To accomplish that, it is necessary to modify the integrable limit for Eq.~\ref{integrablelimit} in 
two ways: firstly, we need to turn on the potential, and secondly we need to consider the dynamics in an FLRW background. Generically, these modifications
break the symmetry under $\phi$ and $t$-translations such that the system is no longer integrable and we need to solve it numerically. 

Notwithstanding, we can talk about the qualitative asymptotic behaviour after some assumptions. Consider for instance the case when $V_\chi =0$ and negligible Hubble friction. In this case, 
\begin{subequations}
\begin{align}
    \dot{p} &= -V_\phi\\
    \dot{\chi} &= \pm \sqrt{2E - p^2f(\chi)^{-1} - V(\phi)},
\end{align}
\end{subequations}
and the $\phi$ evolution does not allow us to reduce the system to one-dimension motion. If $\phi(t)$ is such that $V(\phi(t))$ decreases, we still get a constant $\chi$ velocity asymptotically. If $V(\phi)$ has a minimum, then even if $\phi(t)$ gets at the minimum with significant kinetic energy, the $\dot{\chi}$ dependent damping term will deplete its evolution (note that for $f = e^{\lambda \chi}$, $f_\chi \dot{\chi}/f = \lambda \dot{\chi}$ tends to a constant). So, the presence of a mass term or a cosine potential for $\phi$ does not prevent the energy-transfer mechanism from working.

Finally, let's consider how the Hubble friction term affects the dynamics of the two-field system. This term acts as a time-dependent friction term for both fields and its time dependence breaks the time-translation symmetry. As long as $H(t)$ is large enough for preventing $\phi$ from ``rolling down'' its potential, there can not be efficient energy transfer. However, as soon as the friction is small enough for $\phi$ to start evolving, the energy transfer commences as described before. Then, after most of the kinetic energy is transferred, $\dot{\chi}$ will stop increasing and $\phi$ will again be subjected to the Hubble friction and its potential. On the other hand, in the absence of a potential for $\chi$, $\dot{\chi}$ will asymptote to zero, since the Hubble friction will act to halt the $\chi$ evolution. The potential for $\chi$ may change this late behaviour, but as long as it does not have enough amplitude to play a role during the energy transfer phase, the dynamics during that period goes as described before. Indeed, after sequestering the energy from $\phi$, the energy density in $\chi$ redshifts as a stiff component, so that its potential energy might only be relevant once the energy density is completely negligible. We discuss numerical solutions of the full system in Section \ref{sec:num}.

In this paper we shall consider a model with $f(\chi) = e^{\lambda\chi}$ and a cosine potential for $\phi$, $V = m_\phi^2 (1-\cos \phi)$. Strictly speaking $\lambda$ is an $\mathcal{O}(1)$ number which depends on the choice of internal manifolds for string compactification, however we will henceforth treat it as a free parameter of our theory. Let us now estimate $f_{\rm KMIX}(z_c)$, the fraction of the total energy density in the fields $\phi$ and $\chi$ at the redshift $z_c$ where it reaches its maximum. Assuming the $\chi$-potential does not play a role in the dynamics of our model, we have the following parameters: the initial values of the fields $\phi_i$ and $\chi_i$, the mass $m_\phi$, and the kinetic coupling parameter $\lambda$. Our goal is to now find $f_{\rm KMIX}(z_c)$ and $z_c$ in terms of these parameters.

Similar to EDE, $\phi$ only becomes dynamical when $3H \sim \sqrt{V_{\phi\phi}/f}$. So, $z_c$ is fixed by
\begin{equation}
    3H(z_c) \approx e^{-\lambda \chi_i/2}m_\phi \cos \phi_i.
\end{equation}
Before $z_c$, the fraction of energy in $\chi$ is much smaller than in $\phi$. Moreover, since $\phi$ is essentially constant until approximately $z_c$, it will not source $\chi$ leading up to this point. Thus, $\chi(z_c) \approx \chi_i$ and we can approximate (neglecting the $\chi$ contribution to $V$)
\begin{equation}
    f_{\rm KMIX}(z_c) \approx \frac{e^{\lambda \chi_i}\dot{\phi}(z_c)/2+V(\phi(z_c))}{\rho_\text{tot}(z_c)}.
\end{equation}
Neglecting the kinetic energy in $\phi$ and approximating $\rho_\text{tot}(z_c)$ by $3H^2(z_c) \sim |V_{\phi\phi}(\phi_i)| /3f(\chi_i)$, we find that
\begin{equation}
    f_{\rm KMIX}(z_c) \propto f(\chi_i)\frac{V(\phi_i)}{|V_{\phi\phi}(\phi_i)|}.
\end{equation}
Hence, $f_{\rm KMIX}(z_c)$ is essentially set by the initial values of the fields and $\lambda$, in particular it does not depend on $m_\phi$
\begin{equation}
    f_{\rm KMIX}(z_c)\propto e^{\lambda \chi_i/M_{\text{Pl}}}\frac{1-\cos(\phi_i/M_{\text{Pl}})}{\cos(\phi_i/M_{\text{Pl}})}, 
\end{equation}
where we restored the units. The main difference from the standard EDE results is the presence of the $e^{\lambda \chi_i}$ factors in the equations above.

\subsection{Numerics}\label{sec:num}

\begin{figure}[t]
    \centering
    \includegraphics[width=\linewidth]{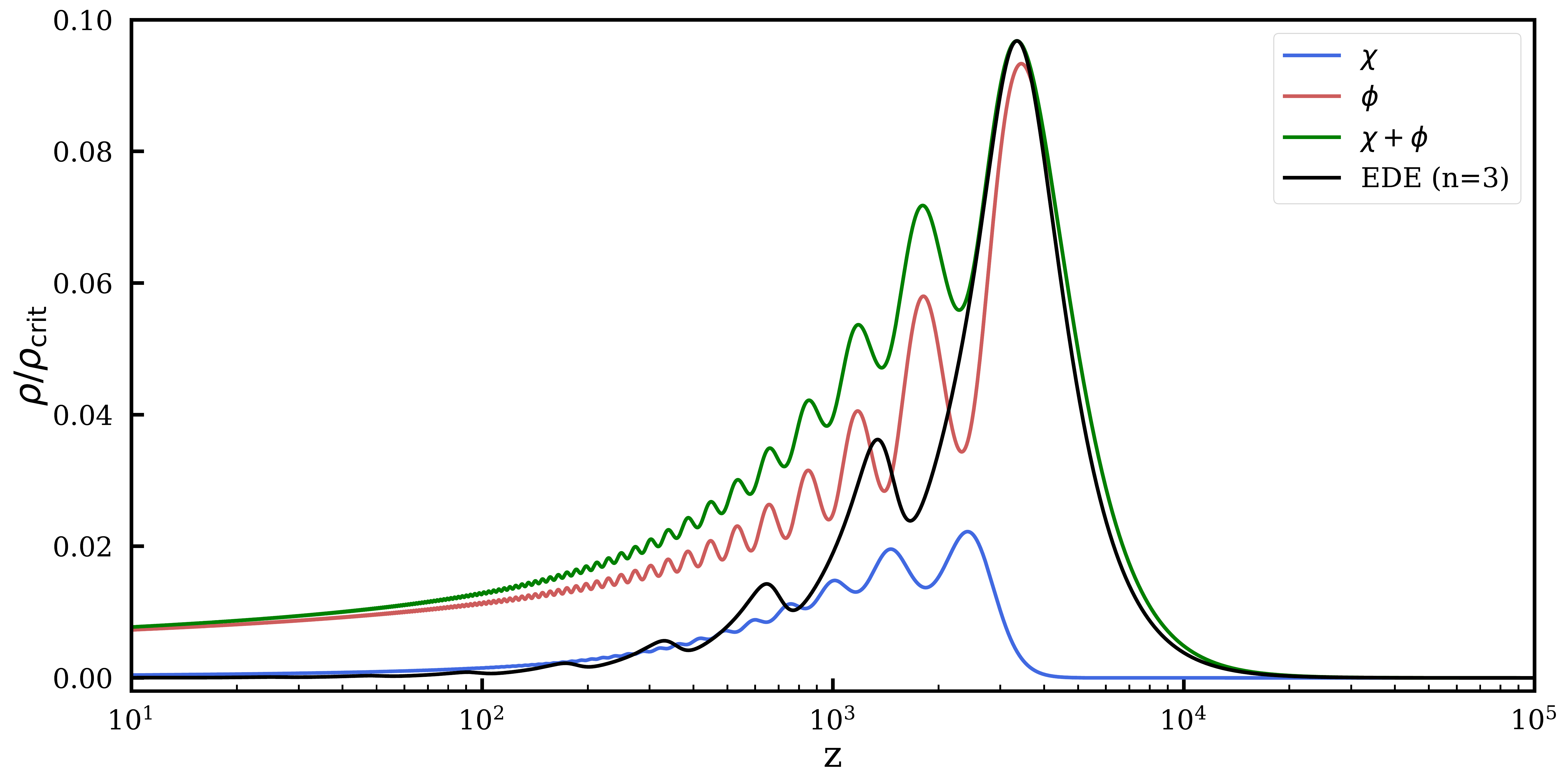}
    \caption{ Evolution of the fraction of the total energy density for the KMIX model with an axion cosine potential (n=1) for $\chi + \phi$ (green), $\chi$ (blue), and $\phi$ (red) with example $H_0$ and $S_8$ tension resolving parameters (see Eq.~\ref{KCparams}). Plotted for comparison is the EDE model (n=3) for the same $f_{\rm EDE}$ and z$_c$ with $\theta_i = 2.83$. }
    \label{fig:density}
\end{figure}

\begin{figure}[t]
    \centering
    \includegraphics[width=\linewidth]{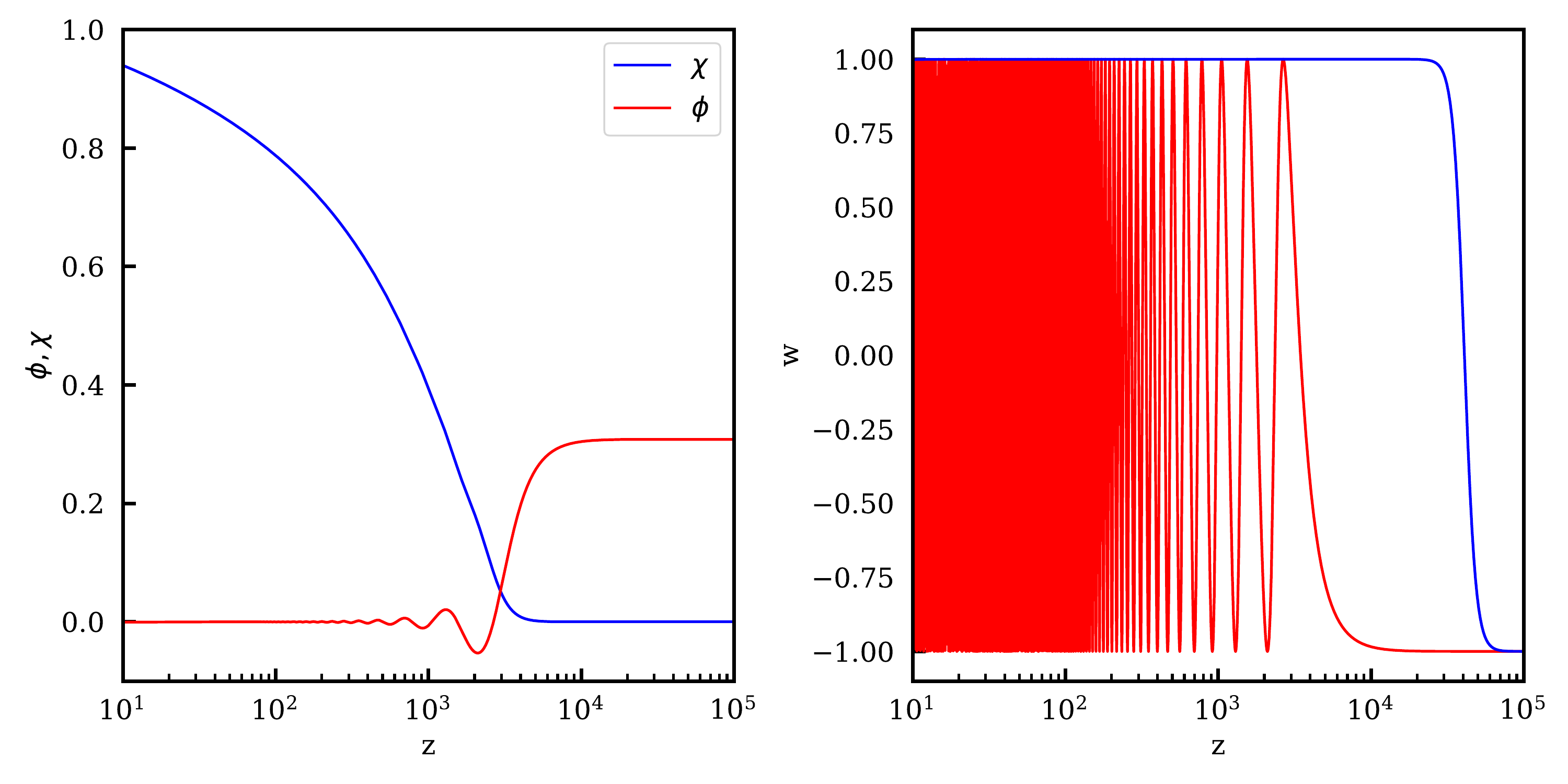}
    \caption{Evolution of the fields for the KMIX model in units of the reduced Planck mass (left panel) and their equations of state (right panel). Cosmology used is the same as that in Fig~\ref{fig:density}.}
    \label{fig:eos}
\end{figure}

\begin{figure}[!t]
    \centering
    \includegraphics[width=\linewidth]{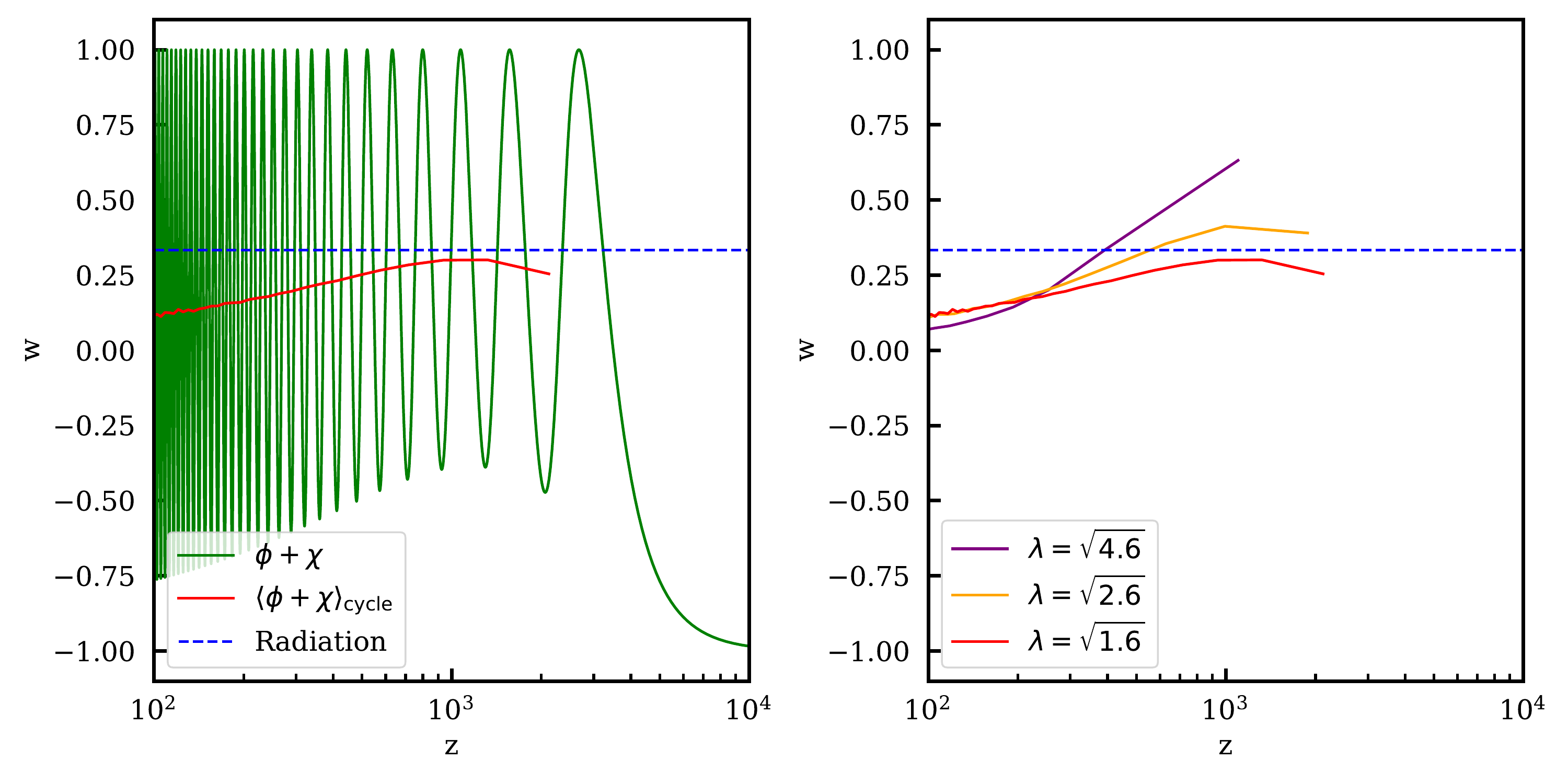}
    \caption{Evolution of the combined equation of state for both fields (green) and the corresponding cycle average (red) for the cosmology used in Fig~\ref{fig:density} (left panel). We also plot the cycle average equations of state for differing values of $\lambda$ (right panel). For reference we plot the equation of state for radiation (blue).}
    \label{fig:average}
\end{figure}

To fully understand the background dynamics we have modified the publicly available Boltzmann code \texttt{CLASS}\footnote{\url{http://class-code.net}} \cite{2011arXiv1104.2932L,2011JCAP...07..034B} to realize our model. For $\phi$ we use a standard axion potential and we supply $\chi$ with a small, constant potential that plays the role of a cosmological constant,
\begin{equation}
    V(\chi,\phi) = m_\phi^2 f_\phi^2 \left(1 - \cos{\frac{\phi}{f_\phi}}\right) + \Lambda.
\end{equation}
The axion mass and decay constant are free parameters, along with the initial value of $\phi$, but $\Lambda$ is chosen such that our cosmology fulfills the budget equation, $\sum_i \Omega_i = 1$. We display the evolution of the fraction of the total energy density in Fig.~\ref{fig:density} for our model with example $H_0$ and $S_8$ resolving parameters (see Eq.~\ref{KCparams}). For comparison, we also overlay the evolution for a $n=3$ EDE model with the same peak energy density, $f_{\rm EDE} = 0.097$, and corresponding redshift, $z_{\rm c} = 3350$, which we calculate using \texttt{CLASS\_EDE}.\footnote{\url{https://github.com/mwt5345/class_ede_v3.2.0}} Generically, EDE models with $n=1$ potentials are ruled out as $w\approx 0$ once the field begins to oscillate at the bottom of its potential, preventing the energy density from depleting fast enough. This results in a significant addition to $\Omega_{\rm cmd}$ and negatively alters the expansion history \cite{Smith:2019ihp}. Fig.~\ref{fig:density} makes clear that this is simply not the case for this model: with natural values for $\lambda$ the increase to $\Omega_{\rm cdm}$ is  $\sim 0.1 {\rm ~to~} 1\%.$ As outlined in the last section, the ability for $\phi$ to redshift sufficiently stems from it being kinetically coupled to $\chi$. This can best be understood by looking at the evolution of the fields and their equations of state, which is presented in Fig.~\ref{fig:eos}. We see that after $\phi$ starts to roll down its potential $\chi$ grows rapidly, which is intuitive as $\dot\phi^2$ sources $\chi$ (see Eq.~\ref{chi_eom}). This realizes a transfer of energy from $\phi$ into $\chi$, with $w\approx1$, that allows the total energy density to redshift from $\approx 1/a^3$ to $1/a^6$ depending on the choice of $\lambda$.
To demonstrate this, we plot in the left panel of Fig.~\ref{fig:average} the combined equation of state for both fields, including the cycle average, for the example presented in Fig.~\ref{fig:density}. This shows that the combined energy density in this case redshifts (approximately) as radiation initially, which is qualitatively similar to $n = 2$ EDE models.  The efficiency of this transfer is controlled primarily by the choice of $\lambda$: larger $\lambda$'s result in faster depletion of the energy density, which can also be understood from the source term in Eq.~\ref{chi_eom}. We show this explicitly in the right panel of Fig.~\ref{fig:average} where the cycle averaged equation of state for different values of $\lambda$ is plotted.

\section{Perturbations}\label{pert}

To understand the role perturbations play in our model we derive the perturbation equations by varying the action, Eq.~\ref{action}, to quadratic order. Thus we will not take a fluid approximation and instead solve the full linearized, coupled Klein-Gordon equations along with the usual equations for the metric perturbations. In synchronous gauge we find that the equation of motion for the perturbation in $\chi$ is given by,
\begin{equation}
    \label{pert1}
    \delta \chi'' + 2 a H \delta \chi' + \left[ k^2 + a^2 V_{\chi\chi} - \frac{1}{2} \lambda^2 f (\phi')^2 \right] \delta\chi = \lambda f \phi' \delta\phi' - \frac{h'\chi'}{2},
\end{equation}
and for $\phi$,
\begin{equation}
    \label{pert2}
    \delta \phi'' + \left[2aH + \lambda \chi' \right]\delta \phi' + \left[k^2 + \frac{a^2}{f}V_{\phi\phi} \right]\delta\phi = \lambda\left(\frac{a^2}{f}V_\phi\delta\chi - \phi'\delta\chi'\right) - \frac{h'\phi'}{2},
\end{equation}
where $'$ denotes a derivative with respect to {\it conformal time},  $H$ is the Hubble parameter in \textit{cosmic time}, and $h$ is the trace of the spatial metric perturbation. In the limit that the initial field value for $\chi$ vanishes, which we generally assume, and the absence of kinetic coupling, $\lambda = 0$, the dynamics for perturbations (and background) reduce to that of EDE. Arguably, this is a desirable feature given the that EDE can achieve great fits to the CMB with non-negligible parameter shifts relative to $\Lambda$CDM.
Indeed, as pointed out in \cite{Lin:2019qug}, any new dark species will modify the phase and amplitude of the CMB peaks from gravitationally driving the photon-baryon acoustic oscillations. In the case of EDE-like models, the scalar field perturbations oscillate leading up to recombination, rather than grow, which results in a suppression of the Weyl potential. While the physical dark matter density can be increased to maintain a good fit to the CMB, this affects the perturbations in the late-time universe by enhancing the growth of structure as quantified by a noticeable increase in the $S_8$ parameter relative to $\Lambda$CDM. Clearly perturbations are what will make or break any model aimed at addressing the $H_0$ tension, so the new dynamics in our model are welcome to confront the challenges faced in the late-time universe.

\subsection{Resolving the large-scale structure tension}

One important feature of our model is that a fraction of $\Omega_{\rm cdm}$ will necessarily be an $\mathcal{O}(10^{-27} {\rm ~eV})$ axion. This is unavoidable as the contribution is from $\phi$ after its has fallen down its potential - a direct byproduct of a true axion potential. The fraction of dark matter in the light axion, $F = \Omega_\phi /\Omega_{\rm cdm}$, is controlled by $\lambda$ - larger values result in smaller $F$. As is well known, the presence of such a light axion can have a major impact on the growth of large-scale structure \cite{Marsh:2010wq,Amendola:2005ad,Allali:2021azp} as the sound speed for scalar dark matter is scale dependent \cite{hu, Hwang:2009js}. Below the Jeans scale, $k_J/ a = 6^{1/4} \sqrt{H m}$, scalar  dark matter density perturbations do not grow, but oscillate, resulting in a suppression of large-scale structure. However, during radiation domination the growth of dark matter density perturbations are already strongly suppressed due to the M\'esz\'aros effect \cite{1974A&A....37..225M}, so deviations from $F = 0$ do not become relevant until after matter-radiation equality. We can get a handle on the expected impact by first calculating the Jeans wave number for $\delta\phi$ at equality,  $k^{\rm eq}_{J}  \sim 10~m_{-22}^{1/2} {\rm~Mpc}^{-1}$, which for a typical axion in our theory has $k^{\rm eq}_{J} \sim 0.03~{\rm Mpc}^{-1}$. The impact of $F \neq 0$ will be a modification of the typical linear growth of dark matter perturbations post-equality for scales below $k^{\rm eq}_J$: $\delta_{\rm cdm} \propto a \rightarrow a^q$. The deviation can be approximated as \cite{Marsh:2010wq, Amendola:2005ad},
\begin{equation}
    q = (\sqrt{25 - 24 F} - 1)/4,
\end{equation}
and the suppression of the matter power spectrum for $k > k^{\rm eq}_J$ is given by,
\begin{equation}
    \frac{\mathcal{P}_{\phi + {\rm cdm}}}{\mathcal{P}_{\rm cdm}} \approx \left(\frac{k^{\rm eq}_J}{k}\right)^{8(1 - q)},
\end{equation}
which manifests as a noticeable step in the matter power spectrum at $k_J^{\rm eq}$ \cite{Marsh:2010wq}. For a typical axion in our theory with $F = 0.01$,  well outside current constraints from the CMB and LSS \cite{Hlozek:2014lca}, there will be a $\sim 10\%$ suppression in power over scales sensitive to $S_8$, more than enough to resolve the tension.

While it is less influential, another feature of our model that helps alleviate the tension is that perturbations should not be suppressed as effectively pre-recombination. We can understand this by looking at the perturbed Klein-Gordon equations, Eqs.~\ref{pert1}~and~\ref{pert2}. The most obvious effect is the presence of the extra source terms which should help to alleviate the decay of the Weyl potential. Note that in the thermal friction model presented in \cite{Berghaus:2022cwf} there is also a source term in the perturbation equations between the scalar field and dark radiation which, in that case, has the consequence of degrading the fit to the CMB. This stems primarily from the extra source term (which is $\propto \dot\phi^2$) dominating over the sourcing of the metric perturbations, resulting in the suppression of the radiation perturbations relative to other species. However, the source term in our model will be a much weaker effect as the transfer of energy between fields has less influence on the background dynamics compared to thermal friction models. Thus, our source terms should serve as a small correction to metric perturbations and, by extension, reduce the required physical dark matter density to fit the CMB.

\subsection{Avoiding Planckian field excursions}

A less prominent, but still important, theoretical hurdle of best-fit EDE models with $n = 3$ potentials is the preference for large initial field displacements, i.e. $\theta_i \sim \pi$. This is problematic as best-fit values for the decay constant are near the Planck scale \cite{Hill:2020osr} meaning the EDE scalar undergoes Planckian field excursions, $\Delta \phi \sim$ $M_{\rm pl}$, which is a major theoretical issue from the perspective of the Distance and the Weak Gravity Conjectures \cite{Ooguri:2006in,Arkani-Hamed:2006emk}. What forces EDE models to prefer large field values is the impact on perturbations, particularly from constraints by Planck EE and TE power spectra.\footnote{While \cite{Hill:2021yec} are not able to well constrain $\theta_i$, for a CMB analysis independent of Planck with ACT data there appears to be a preference for much smaller values of $\theta_i$.} Specifically, as pointed out in \cite{Lin:2019qug,Smith:2019ihp}, for a good fit to the CMB it is important to maximize the number of modes around the time of energy injection with a sound speed $c_s^2 \approx w$, provided $w > 1/3$. The range of modes inside the horizon with the right sound speed is maximized (see \cite{Smith:2019ihp} for more details) for $\theta_i^{n - 1} /\sqrt{|E_{n,\theta\theta}|} \gg 1$, where $E_{n} = (1 - \cos{\theta})^n$ is the effective potential. For $n = 3$ this requires $\theta_i \sim \pi$, whereas this ratio becomes large for $\theta_i \sim \pi/2$ in $n = 1$ models. This can be understood as the $n=1$ potential becoming convex for smaller $\theta_i$ relative to $n = 3$ potentials. With this insight in mind, our model should accommodate the same energy injection as $n = 3$ EDE but with $\Delta \phi \sim 0.35~M_{\rm pl}$, alleviating the concern for Planckian field excursions.

\subsection{Parameter space of the model}

\begin{figure}[!t]
    \centering
    \includegraphics[width=\linewidth]{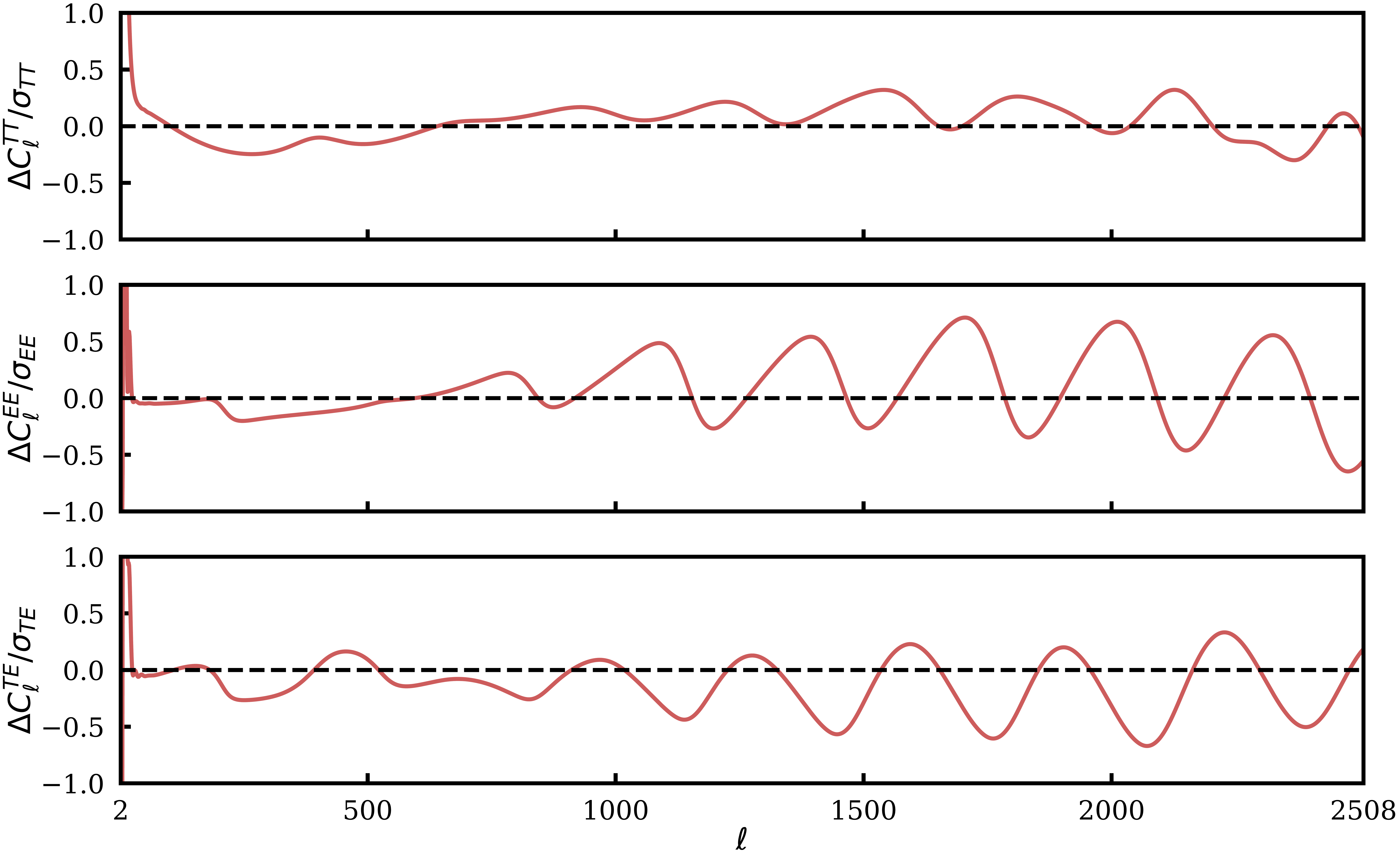}
    \caption{Difference in CMB temperature (top), polarization (middle), and cross-correlation (bottom) power spectra between best-fit $\Lambda$CDM model from the Planck 2018 analysis and KMIX with parameters given in Eq.~\ref{KCparams} in units of cosmic variance per $\ell$-mode of the best-fit $\Lambda$CDM model.}
    \label{fig:cmb}
\end{figure}

\begin{figure}[t]
    \centering
    \includegraphics[width=\linewidth]{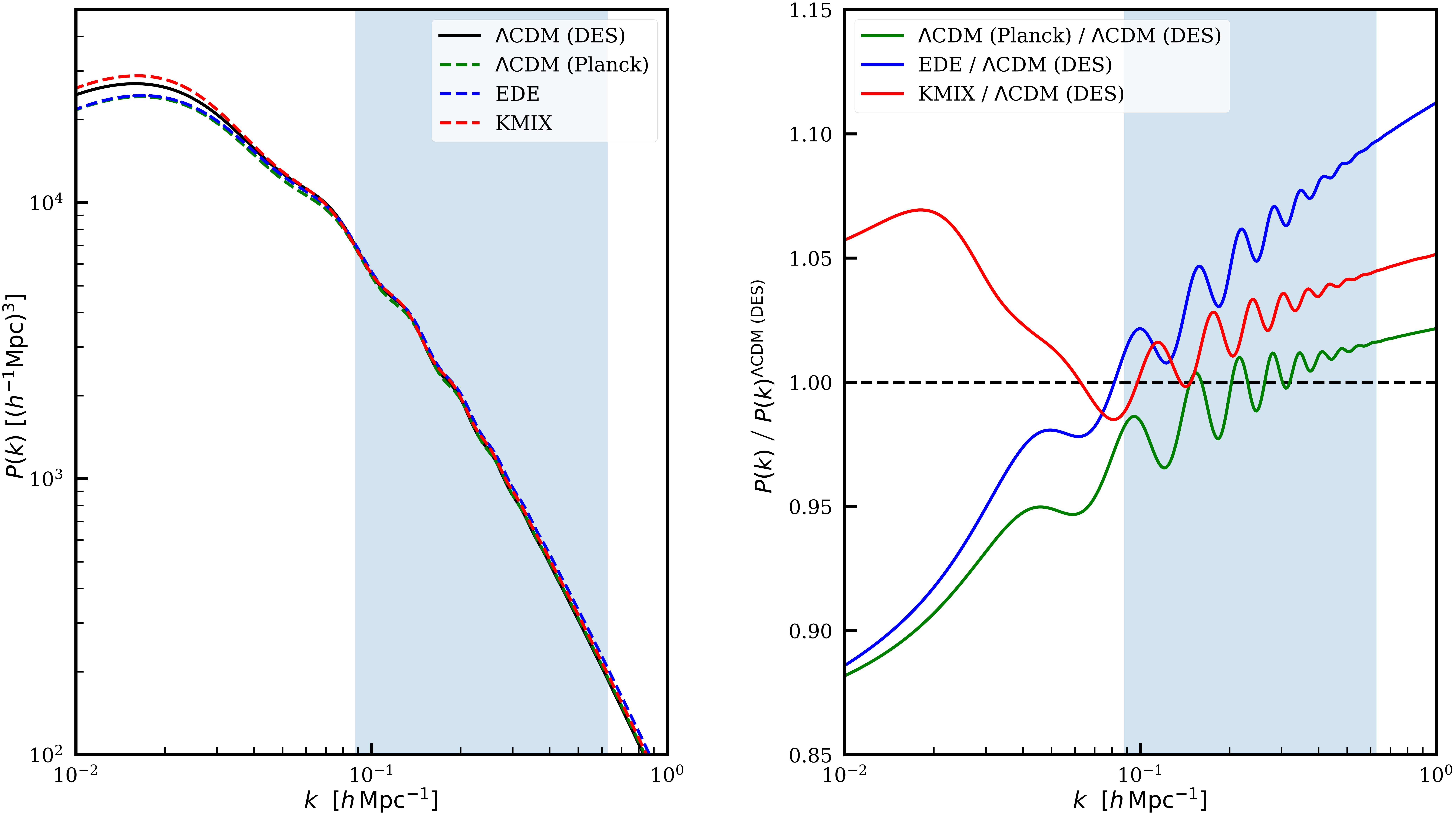}
    \caption{Linear matter power spectra (left panel) and ratio of spectra (right panel) at $z = 0$ for $\Lambda$CDM with the DES-Y3 + BAO + BBN best-fit values $\Omega_m = 0.296$, $S_8 = 0.809$, $\Omega_b = 0.048$, H$_0 = 67.3$ km/s/Mpc from \cite{DES:2021wwk} (black), the best-fit $\Lambda$CDM from the Planck 2018 analysis \cite{Planck:2018vyg} (green), EDE best-fit from \cite{Hill:2020osr} (blue), and for the KMIX model with parameters given in Eq.~\ref{KCparams} (red). Ratios in the right panel are taken with respect to the DES-Y3 $\Lambda$CDM spectrum, thus giving an indication of how well the other models’ predictions match the DES-Y3 constraints. The blue shaded region shows the approximate range of comoving wavenumbers probed in the DES-Y3 analysis.}
    \label{fig:mPkDES}
\end{figure}

To see if our theory does indeed exhibit the right properties to address the Hubble and LSS tensions, we have explored the parameters space of our model. In Fig. \ref{fig:cmb} we show the CMB temperature, polarization, and cross-correlation power spectra for our model with example $H_0$ and $S_8$ resolving parameters,
\begin{align}
     \label{KCparams}
     H_0&=72.75 \, {\rm km/s/Mpc} &   100\omega_b&=2.283, \\
     \omega_{\rm cdm}&=0.1285 &10^9 A_s &= 2.250, \nonumber \\
     n_s&=0.970  &    \tau_{\rm reio}&=0.055, \nonumber \\ 
     \lambda &= \sqrt{1.6} & \theta_i &= 1.60,\nonumber \\
     m_\phi &=1.00\times10^{-27}\,{\rm eV} & f_\phi &= 4.50\times10^{26}\,{\rm eV},\nonumber
\end{align}
relative to $\Lambda$CDM with best-fit parameters from the Planck 2018 analysis \cite{Planck:2018vyg},
\begin{align}
     \label{smithparamsLCDM}
     H_0&=67.32 \, {\rm km/s/Mpc} &   100\omega_b&=2.283, \\
     \omega_{\rm cdm}&=0.1201 &10^9 A_s &= 2.210, \nonumber \\  n_s&=0.9661 &    \tau_{\rm reio}&=0.0543, \nonumber 
\end{align}
in units of cosmic variance per $\ell$-mode of the best-fit $\Lambda$CDM model. The residuals between the two models are well within cosmic variance, so the parameters given by Eq.~\ref{KCparams} can be understood to represent a good realization of the CMB. Interestingly, considering the values used in Eq.~\ref{KCparams}, we can already see the impact of the source term in the perturbations. The best-fit Planck EDE model from \cite{Smith:2019ihp} necessitated $\omega_{\rm cdm} = 0.1306$, an increase of $\sim 9\%$ over $\Lambda$CDM, to achieve an $H_0 = 72.19$ km/s/Mpc. In contrast, our example accommodates a larger $H_0$ value but with a \textit{lower} physical dark matter density.

To see how perturbations for this example evolve into the late-time universe we also calculate the linear matter power spectrum
in \texttt{CLASS}, Fig.~\ref{fig:mPkDES}.
For comparison we also plot the power spectra for the best-fit $\Lambda$CDM model from the Planck 2018 analysis \cite{Planck:2018vyg}, best-fit Planck EDE for CMB with parameters from \cite{Hill:2020osr}, and for $\Lambda$CDM with the best-fit DES-Y3 + BAO + BBN values from \cite{DES:2021wwk}. The blue region corresponds to the approximate comoving wave numbers probed by DES. What is clear from this plot is that our model does not suffer from the same excess of power present in the best-fit Planck $n= 3$ EDE model and, instead, performs similarly to the best-fit Planck $\Lambda$CDM model. In fact, this is exactly what we expect to see for our model. Specifically, notice that the matter power spectrum has a clear step in the residuals plot starting  $\sim 0.03$ Mpc$^{-1}$ and realizes a suppression of $\sim~10\%$. This is exactly the effect we anticipate for the axion in our theory if $F = 0.01$ (see Fig.~\ref{fig:density}). For this example we calculate $S_8 = 0.807$, whereas the best-fit Planck $\Lambda$CDM model has $S_8 = 0.831$ and the best-fit Planck EDE model has $S_8 = 0.847$. Note that the $S_8$ value for our model is lower than Planck, despite having more power in Fig.~\ref{fig:mPkDES}, as $\Omega_m$ is smaller. While Eq.~\ref{KCparams} are not the best-fit parameters for our model, this would of course necessitate a full MCMC analysis, these results corroborate our expectation from the theory that there should be regions of parameter space that have the right dynamics to solve \textit{both} the Hubble and $S_8$ tensions.  Concretely, for values that produce CMB power spectra close to the best-fit $\Lambda$CDM result but with large $H_0$ values, we see in the late-time universe the matter power spectrum does not have a significant excess of power at scales sensitive to $S_8$.

\section{Discussion and Conclusion}\label{DnC}

In this work we have presented a new string theory inspired model which has the ability to resolve both the Hubble and $S_8$ tensions. The dynamics are governed by two fields, an axion and a dilaton which are kinetically coupled. Our axion plays a similar role to the scalar field in EDE in that its purpose is to decrease the size of the sound horizon. The dilaton plays a critical role through its coupling to the axion: it siphons off energy density as the axion falls down its potential allowing it to redshift faster. These dynamics allow use of a standard axion cosine potential viable for solving the Hubble tension, unlike in EDE. From a theoretical perspective, this a significant improvement over EDE models where one must choose a potential with fine-tuned coefficients for leading terms in the instanton expansion to get the correct background dynamics.

More still, the axion in our model plays a double role. Given that it must have a mass $\mathcal{O}(10^{-27}{\rm~eV})$ to resolve the Hubble tension and contribute a small fraction to dark matter post-recombination, it will naturally suppress power on scales sensitive to $S_8$. Concretely, if it contributes $1\%$ to $\Omega_{\rm cdm}$  the matter power spectrum will be suppressed for relevant $k$ by $\sim 10\%$, more than enough to resolve the $S_8$ tension. While such a small fraction is not ruled out by the CMB and LSS \cite{Hlozek:2014lca}, future measurements of the 21 cm signal \cite{Flitter:2022pzf} will be able to constrain small fractions of very light axion dark matter. Interestingly, in \cite{Allali:2021azp} a phenomenological model was studied with a period of early dark energy but supplemented by a small contribution from a separate very light scalar. Their results were found to alleviate both the $H_0$ and $S_8$ tensions. Since our model can be seen as a concrete and physically well motivated realization of their toy model, this makes the prospect for viability of our model promising. Furthermore, a similar action to Eq.~\ref{action} was studied in \cite{Allali:2022yvx}, but in the context of axion dark matter relic abundances. The dynamics explored in that work are similar to how $F$ can be controlled with a choice of $\lambda$ in our model.

From the theory side, it is worthwhile to recall that the amplitude of the axion potential should be a fraction of the energy scale of the background close to  recombination. Since our model is likely to be embedded into string theory, this implies that we should be able to reproduce $\mathcal{O}(1)$ eV-scale physics from the string scale. Assuming the axionic Weak Gravity Conjecture, we need an ultralight axion for that to be possible. So, one might think that there is some sort of ``coincidence'' or ``why now'' issue with our construction, like in all EDE models. However, this is likely to be solved by the UV completion in our model while, on the other hand, there are no proposals for such completions in the $n=3$ EDE case. In fact, semi-realistic string models containing ultralight axions were recently discussed in \cite{ Cicoli:2021gss,Bernardo:2022ztc}. 

Going beyond analytics, we have also used a modified version of the Boltzmann code \texttt{CLASS} to explore the parameter space of our model. Calculating the background dynamics, CMB power spectra and linear matter power spectrum we have identified regions that can accommodate larger values for $H_0$ without increasing $S_8$ -- a $\sim 10\%$ increase in $H_0$ from the Planck inferred value results in $S_8$ values consistent with LSS constraints. However, we again stress that we have not reported the best-fit parameters for our theory but have instead displayed a representative cosmology that exhibits the features of a Hubble and $S_8$ tension resolving model. We leave the detailed investigation of the perturbations and a full MCMC analysis for future work.

Besides what we have already discussed, one of the appealing properties of our model is that we only have one additional free parameter compared to EDE, this is $\lambda$. What makes this parameter particularly interesting is that it is related to internal manifolds of the string compactification, realizing a potential avenue to constrain string theory. From another point of view, similar to how one makes a choice of $n = 3$ in EDE models, one could also select a well informed $\lambda$ and reduce the model parameters by one. Furthermore,  a simple $m^2\phi^2$ potential will suffice to resolve the Hubble tension at the background level, i.e. just the leading term in the limit $\phi/f_\phi$ is small, with the added benefit of further reducing the number of parameters. However, its not clear that data will favor this choice as power law potentials generally do not fit CMB data well. 

To conclude on a more speculative note, it is interesting that the dynamics of our model prefer that the potential for the dilaton be small. While our choice to implement (late-time) dark energy as the constant potential of $\chi$ was done to fulfill the budget equation in \texttt{CLASS}, it would be interesting if a connection between a period of early dark energy and dark energy could be made. Indeed, it is well known that a dilaton can explain late-time dark energy (see for example \cite{Burgess:2021obw}) and motivates exploring a true dilaton potential, $V(\chi) = V_0e^{\beta \chi}$.

\section*{Acknowledgments}
We would like to thank Kim Berghaus, Robert Brandenberger,  Colin Hill, and Evan McDonough for helpful comments on an early draft of this work. We additionally thank Steven Clark, Mikhail Ivanov, Tanvi Karwal, Oliver Philcox,  and David Spergel for helpful discussions. H.B. would like to thank the CCA/Flatiron Institute, the Symmetry and Cosmology group of the Humboldt University of Berlin and ETH-Zurich for hospitality during the execution of this work. H.B. research is supported by the Fonds de Resercher du Qu\'ebec (PBEEE/303549) and partially by funds from NSERC.

\appendix

\section{String Theoretic Embedding of KMIX}\label{string-append}

We now review how scalar sectors of the form in Eq.~\ref{scalarsectoraction} arise in string theory. It is customary to compactify string theory on internal spaces that preserve $\mathcal{N}=1$ at low energies \cite{Grana:2005jc,Blumenhagen:2006ci}. After dimensional reduction, the four-dimensional action contains many fields associated to the shape and size of the internal manifold, but since the effective theory should be a four-dimensional $\mathcal{N}=1$ supergravity, these should come in supersymmetric multiplets whose action is fixed by a few functions of the multiplets: a K\"ahler potential $K$ that fixes the kinetic term of the scalar moduli, a gauge kinetic function that fixes the kinetic term of gauge fields and a superpotential that determines the interactions among the fields \cite{wess2020supersymmetry,freedman2012supergravity}. An important issue in string phenomenology is to explain how to obtain these functions from ten-dimensional physics, and the literature on moduli stabilization is very extensive. For us, the important feature of these effective actions is the presence of the kinetic coupling mentioned above, so we will focus on the K\"ahler potentials that originates from string theory.

For both heterotic and type II superstring theories, the K\"ahler potential for certain chiral multiplets has a ``no-scale'' form \cite{Lahanas:1986uc}, generically given by \cite{Cicoli:2013rwa}
\begin{equation}\label{Kaehlerpotential}
    K = - \ln (S+ \Bar{S}) - \ln \mathcal{V},
\end{equation}
where the so called axio-dilaton moduli is a complex field whose real part is related to the dilaton $\Phi$ and its imaginary part is an axion $a$, $S = e^{-\Phi} +ia$. The axion appears after dualizing the four-dimensional components of the Kalb-Rammond 2-form field that is present in any perturbative limit of the theory. For that reason, $a$ is also referred to as the model-independent axion (because it does not depend on the details of the compactification) \cite{Svrcek:2006yi}. The quantity $\mathcal{V}$ is the volume of the internal manifold in string-scale units and, for Calabi-Yau (CY) three-folds,  
\begin{equation}
    \mathcal{V} = \frac{1}{6}d_{ijk}t^i t^j t^k,
\end{equation}
where $t^i = (T^i+\overline{T}^i)/2$ with $T^i$ being the (complexified) size of 2-cycles in the internal geometry\footnote{This is specific to hetoritic strings, for type II strings one usually writes $\mathcal{V}$ in terms of 4-cycles moduli. This technical point is irrelevant to our discussion.} and $d_{ijk}$ is the triple intersection number of the CY manifold.

The K\"ahler metric $K_{I\bar{J}} = \partial_I \partial_{\bar{J}}K$ is the field space metric that sets the kinetic terms for the complex fields $S$ and $T^i$ (the indices $I$ and $J$ are associated to axio-dilaton and K\"ahler fields). The moduli fields are canonically normalized only if $K_{I\bar{J}}$ can be diagonalized everywhere 
in the field space. Generically, for K\"ahler potentials of the form in Eq.~\ref{Kaehlerpotential}, there will be a kinetic coupling between pairs of moduli fields. As an illustrative example, for a straight toroidal orbifold where all direction have the same radius, there is only one modulus, setting the overall volume of the torus.
In this case, we might write $\mathcal{V} = (T + \overline{T})^3$ \cite{Polchinski:1998rr}. Then, the kinetic term for the axio-dilaton and K\"ahler moduli will be
\begin{align}
K_{I\bar{J}}\partial_\mu T^I \partial_\nu \overline{T}^J g^{\mu\nu} &= \frac{1}{(S+\overline{S})^2}\partial_\mu S \partial^\mu \overline{S} + \frac{3}{(T+\overline{T})^2}\partial_\mu T \partial^\mu \overline{T}\nonumber\\
&= \frac{1}{4}(\partial \Phi)^2 + \frac{e^{2\Phi}}{4}(\partial a)^2 + \frac{3}{4}(\partial \Psi)^2 + \frac{3e^{-2\Psi}}{4}(\partial b)^2,
\end{align} 
where we decomposed $S = e^{-\Phi} +ia$ and $T = e^{\Psi} + ib$ in the second equality. Hence, the kinetic coupling function has an exponential form. We should stress that such mixing is mainly due to the
existence of extra dimensions and the kinetic term for $p$-form fields in higher dimensions, so it might also appear in other extra dimensional theories whether they come from string theory or not. For more complicated internal manifolds, the kinetic coupling might be even more involved, but since $K_{i\bar{j}}$ is hermitian, it will not depend on the imaginary part of $T^i$ and so its kinetic term will always be non-canonical and couple with $(T+\overline{T})^i$. That is, string theory axions generically exhibit the kinetic coupling with other moduli. Of course, if the latter are assumed to be stabilized, the axion can be canonically normalized and its decay constant will depend on these stabilized values (see \cite{Arvanitaki:2009fg,Cicoli:2012sz} for early studies on string axions in cosmology). But if that is not the case, the axion will have a non-trivial kinetic coupling or, in other words, a spacetime-dependent decay constant. Varying decay constants may also appear for throat-axions, which are axions that descend from fluxes along warped throats in the internal space \cite{Dasgupta:2008hb, Franco:2014hsa,McDonough:2018xzh,Hebecker:2018yxs, Cicoli:2021gss}.

\section{Perturbation Equations}\label{pert-append}

In this appendix we derive the perturbations for our theory and discuss the implementation of their dynamics in \texttt{CLASS}.\footnote{Note that until \texttt{CLASS} v2.10 differential equations were solved with respect to conformal time. This forced one to use a fluid approach due to the increasing frequency of scalar field oscillations with $\tau$ for such light scalar fields with $n = 1$ potentials. This problem is now alleviated as new versions of \texttt{CLASS} solve the dynamics with respect to $\log{a}$.} We start by decomposing the two scalar fields into smooth time dependant background and spatially varying perturbations,
\begin{equation}
    \phi(\vec{x},\tau) = \phi(\tau) + \delta \phi(\vec{x},\tau),
\end{equation}
\begin{equation}
    \chi(\vec{x},\tau) = \chi(\tau) + \delta \chi(\vec{x},\tau).
\end{equation}
In synchronous gauge the perturbed metric is given by,
\begin{equation}
    ds^2 = a^2(\tau)\left(-d\tau^2 + (\delta_{ij} + h_{ij})dx^idx^j\right).
\end{equation}
The metric perturbation $h_{ij}$ can be decomposed into trace and traceless components,
\begin{equation}
    h_{ij} = \frac{1}{3} h \delta_{ij} + \left(\partial_i\partial_j - \frac{1}{3}\delta_{ij}\nabla^2\right)\mu + \partial_{(i} A_{j)} + h^{\rm T}_{ij},
\end{equation}
where $h=h^i_i$ and $\mu$ are scalar, $A_i$ vector, and $h^T_{ij}$ tensor perturbations, respectively. Considering only scalar perturbations, the real space degrees of freedom are related to those in k-space as, 
\begin{equation}
    h_{ij}(\Vec{x},\tau) = \int d^3k e^{i\Vec{k}\cdot\Vec{x}}\left[ \hat{k}_i\hat{k}_j h(\vec{k},\tau) + \left(\hat{k}_i\hat{k}_j - \frac{1}{3}\delta_{ij} \right) 6\eta(\vec{k},\tau) \right],
\end{equation}
where $\vec{k} = k \hat{k}$. In Fourier transforming we use $\partial_i \rightarrow i k$ and $\partial_i\partial^i \rightarrow -k^2$. The equations of motion are found from varying the action, Eq. \ref{action}, to quadratic order in $\delta \chi$ and $\delta \phi$. Doing this we find,
\begin{equation}
    \delta \chi'' + 2 a H \delta \chi' + \left[ k^2 + a^2 V_{\chi\chi} - \frac{1}{2} \lambda^2 f (\phi')^2 \right] \delta\chi = \lambda f \phi' \delta\phi' - \frac{h'\chi'}{2},
\end{equation}
\begin{equation}
    \delta \phi'' + \left[2aH + \lambda \chi' \right]\delta \phi' + \left[k^2 + \frac{a^2}{f}V_{\phi\phi} \right]\delta\phi = \lambda\left(\frac{a^2}{f}V_\phi\delta\chi - \phi'\delta\chi'\right) - \frac{h'\phi'}{2}.
\end{equation}
To include our model in \texttt{CLASS} we need to translate our second order perturbed Klein-Gordon equations into the equivalent set of first order fluid equations. The energy-momentum tensor for our model is just the sum of the contributions from both fields,
\begin{equation}
    T_{\mu\nu} =  T^\phi_{\mu\nu} +  T^\chi_{\mu\nu},
\end{equation}
where the energy-momentum tensor for $\chi$ is the same as for a typical scalar field,
\begin{equation}
    T^{(\chi) \mu}_{~~~~~\nu} = \partial^\mu \chi \partial_\nu \chi - \delta^\mu_\nu \left( \frac{1}{2} \partial^\alpha \chi \partial_\alpha \chi + V(\chi) \right).
\end{equation}
The energy-momentum tensor for $\phi$ is modified due to the kinetic coupling and becomes,
\begin{equation}
    T^{(\phi) \mu}_{~~~~~\nu} = f \partial^\mu \phi \partial_\nu \phi - \delta^\mu_{~~\nu} \left( \frac{1}{2} f \partial^\alpha \phi \partial_\alpha \phi + V(\phi) \right).
\end{equation}
Importantly, the Bianchi identities demand that the combined energy-momentum tensor be conserved,
\begin{equation}
   \nabla_\mu \left(T^{(\phi) \mu}_{~~~~~\nu} + T^{(\chi) \mu}_{~~~~~\nu} \right) = 0.
\end{equation}
Furthermore, we can read off the density and pressure from the energy-momentum tensors,
\begin{equation}
        \rho_\chi = \frac{1}{2a^2}\chi'^2 + V(\chi),
\end{equation}
\begin{equation}
        p_\chi = \frac{1}{2a^2}\chi'^2 - V(\chi),
\end{equation}
which resemble that for a standard scalar field. $\phi$ is again modified due to the kinetic coupling, 
\begin{equation}
        \rho_\phi = \frac{1}{2a^2}f\phi'^2 + V(\phi),
\end{equation}
\begin{equation}
        p_\phi = \frac{1}{2a^2}f \phi'^2 - V(\phi).
\end{equation}

We now want to calculate the perturbed density and pressure for both fields. Starting first with the $\chi$ field,
\begin{equation}
    p_\chi = - \frac{1}{2} g^{\alpha\beta} \partial_\alpha \chi \partial_\beta \chi - V(\chi),
\end{equation}
\begin{equation}
    \rho_\chi = - \frac{1}{2} g^{\alpha\beta} \partial_\alpha \chi \partial_\beta \chi + V(\chi),
\end{equation}
which we perturb to find,
\begin{equation}
    \delta p_\chi = - \frac{1}{2} \delta g^{\alpha\beta} \partial_\alpha \chi \partial_\beta \chi - g^{\alpha\beta} \partial_\alpha \chi \partial_\beta \delta \chi - V_\chi \delta \chi,
\end{equation}
\begin{equation}
    \delta \rho_\chi = - \frac{1}{2} \delta g^{\alpha\beta} \partial_\alpha \chi \partial_\beta \chi - g^{\alpha\beta} \partial_\alpha \chi \partial_\beta \delta \chi + V_\chi \delta \chi.
\end{equation}
Now plugging in for our metric we have,
\begin{equation}
    \delta p_\chi = \frac{1}{a^2}\dot\chi \dot{\delta \chi} - V_{\chi}\delta\chi,
\end{equation}
\begin{equation}
    \delta \rho_\chi = \frac{1}{a^2}\dot\chi \dot{\delta \chi} + V_{\chi}\delta\chi,
\end{equation}
which agree with the general result for a scalar field as derived in  Eq.~A4 of \cite{hu}. Repeating the same procedure but for $\phi$,
\begin{equation}
    p_\phi = - \frac{1}{2} f g^{\alpha\beta} \partial_\alpha \phi \partial_\beta \phi - V(\phi),
\end{equation}
\begin{equation}
    \rho_\phi = - \frac{1}{2} f g^{\alpha\beta} \partial_\alpha \phi \partial_\beta \phi + V(\phi).
\end{equation}
We now consider perturbations to the above,
\begin{equation}
    \delta p_\phi = - \frac{1}{2} f'\delta \chi g^{\alpha\beta} \partial_\alpha \phi \partial_\beta \phi - \frac{1}{2} f \delta g^{\alpha\beta} \partial_\alpha \phi \partial_\beta \phi - f g^{\alpha\beta} \partial_\alpha \phi \partial_\beta \delta \phi - V_\phi \delta \phi,
\end{equation}
\begin{equation}
    \delta \rho_\phi = - \frac{1}{2} f'\delta \chi g^{\alpha\beta} \partial_\alpha \phi \partial_\beta \phi - \frac{1}{2} f \delta g^{\alpha\beta} \partial_\alpha \phi \partial_\beta \phi - f g^{\alpha\beta} \partial_\alpha \phi \partial_\beta \delta \phi + V_\phi \delta \phi.
\end{equation}
These reduce to,
\begin{equation}
    \delta p_\phi = \frac{1}{a^2} f \dot\phi \dot{\delta \phi} - V_\phi \delta \phi +  \frac{\lambda}{2 a^2} f \dot\phi^2 \delta \chi,
\end{equation}
\begin{equation}
    \delta \rho_\phi = \frac{1}{a^2} f \dot\phi \dot{\delta \phi} + V_\phi \delta \phi +  \frac{\lambda}{2 a^2} f \dot\phi^2 \delta \chi,
\end{equation}
and both reduce to the same form as the dilaton in absence of kinetic coupling, i.e. $f=1$.

We now calculate the scalar field velocity perturbations. A fluid with small coordinate velocity $v^i$ has, to linear order in perturbations, a stress tensor component defined as,
\begin{equation}
    \left(\rho + p\right)v_j =\delta T^0_j.
\end{equation}
Using this equation, we will be able to calculate the velocity divergence, $\theta_A \equiv \partial_i v^i_A$, which is needed to solve the linearized Einstein equations in \texttt{CLASS} (see Eq. 22 in \cite{Ma:1995ey}). Starting with $\chi$, we perturb the energy-momentum tensor,
\begin{equation}
    \delta T^{(\chi) \mu}_{~~~~~\nu} = \delta \left[\partial^\mu \chi \partial_\nu \chi - \delta^\mu_\nu \left( \frac{1}{2} \partial^\alpha \chi \partial_\alpha \chi + V(\chi) \right)\right],
\end{equation}
notice that since we want an off diagonal component, we can ignore terms proportional to $\delta^\mu_\nu$ completely. Now we have,
\begin{equation}
    \delta T^{(\chi) \mu}_{~~~~~\nu} = \delta g^{\rho \mu } \partial_\rho \chi \partial_\nu \chi + g^{\rho\mu}\left( \partial_\rho \delta\chi \partial_\nu \chi + \partial_\rho \chi \partial_\nu \delta\chi\right).
\end{equation}
Since we need the $\delta T^0_{\;\;j}$ component, this leaves us with,
\begin{equation}
     \delta T^{(\chi) 0}_{~~~~~j} =  -\frac{1}{a^2} \dot\chi \partial_j \delta\chi,
\end{equation}
We can now plug this into our defining equation for the velocity perturbations,
\begin{equation}
    \left(\rho_\chi + p_\chi\right)v_j =  -\frac{1}{a^2} \dot\chi \partial_j \delta\chi,
\end{equation}
and take the gradient,
\begin{equation}
    \left(\rho_\chi + p_\chi\right)\partial^jv_j =  -\frac{1}{a^2} \dot\chi \partial^j\partial_j \delta\chi.
\end{equation}
Since the velocity divergence is defined as, $\theta_A \equiv \partial_i v^i_A$, and we use the convention that $\partial_j\partial^j = -k^2$, we arrive at our desired result,
\begin{equation}
    \left(\rho_\chi + p_\chi\right)\theta_\chi = \frac{k^2}{a^2}\dot\chi \delta\chi,
\end{equation}
which is the standard result for the velocity divergence of a scalar field, see Eq.~A6 of \cite{hu} and Eq.~16 of \cite{Hlozek:2014lca}.

Let us now consider the contribution for $\phi$. We begin in the same fashion by perturbing the energy-momentum tensor,
\begin{equation}
    \delta T^{(\phi) \mu}_{~~~~~\nu} = \delta \left[ f \partial^\mu \phi \partial_\nu \phi - \delta^\mu_{~~\nu} \left( \frac{1}{2} f \partial^\alpha \phi \partial_\alpha \phi + V(\phi) \right) \right],
\end{equation}
which reduces to,
\begin{equation}
    \delta T^{(\phi) 0}_{~~~~~j} = -\frac{f}{a^2} \dot\phi \partial_j \delta \phi,
\end{equation}
such that we can now write,
\begin{equation}
      \left(\rho_\phi + p_\phi\right)v_j  = -\frac{f}{a^2} \dot\phi \partial_j \delta \phi.
\end{equation}
Taking the divergence of both sides yields,
\begin{equation}
      \left(\rho_\phi + p_\phi\right)\partial^jv_j  = -\frac{f}{a^2} \dot\phi \partial^j \partial_j \delta \phi,
\end{equation}
which gives us,
\begin{equation}
      \left(\rho_\phi + p_\phi\right)\theta_\phi  = \frac{k^2f}{a^2} \dot\phi \delta \phi,
\end{equation}
which again reduces to the standard result for $f=1$.

\bibliographystyle{bibstyle} 
\bibliography{References}

\providecommand{\href}[2]{#2}\begingroup\raggedright\begin{thebibliography}{10}

\bibitem{Verde:2019ivm}
L.~Verde, T.~Treu and A.~G. Riess, \emph{{Tensions between the Early and the
  Late Universe}},
  \href{https://doi.org/10.1038/s41550-019-0902-0}{\emph{Nature Astron.}
  {\bfseries 3} (7, 2019) 891},
  [\href{https://arxiv.org/abs/1907.10625}{{\ttfamily 1907.10625}}].

\bibitem{Planck:2018vyg}
{\scshape Planck} collaboration, N.~Aghanim et~al., \emph{{Planck 2018 results.
  VI. Cosmological parameters}},
  \href{https://doi.org/10.1051/0004-6361/201833910}{\emph{Astron. Astrophys.}
  {\bfseries 641} (2020) A6},
  [\href{https://arxiv.org/abs/1807.06209}{{\ttfamily 1807.06209}}].

\bibitem{Riess:2019cxk}
A.~G. Riess, S.~Casertano, W.~Yuan, L.~M. Macri and D.~Scolnic, \emph{{Large
  Magellanic Cloud Cepheid Standards Provide a 1\% Foundation for the
  Determination of the Hubble Constant and Stronger Evidence for Physics beyond
  $\Lambda$CDM}},
  \href{https://doi.org/10.3847/1538-4357/ab1422}{\emph{Astrophys. J.}
  {\bfseries 876} (2019) 85},
  [\href{https://arxiv.org/abs/1903.07603}{{\ttfamily 1903.07603}}].

\bibitem{Abdalla:2022yfr}
E.~Abdalla et~al., \emph{{Cosmology intertwined: A review of the particle
  physics, astrophysics, and cosmology associated with the cosmological
  tensions and anomalies}},
  \href{https://doi.org/10.1016/j.jheap.2022.04.002}{\emph{JHEAp} {\bfseries
  34} (2022) 49--211}, [\href{https://arxiv.org/abs/2203.06142}{{\ttfamily
  2203.06142}}].

\bibitem{DiValentino:2021izs}
E.~Di~Valentino, O.~Mena, S.~Pan, L.~Visinelli, W.~Yang, A.~Melchiorri et~al.,
  \emph{{In the realm of the Hubble tension\textemdash{}a review of
  solutions}}, \href{https://doi.org/10.1088/1361-6382/ac086d}{\emph{Class.
  Quant. Grav.} {\bfseries 38} (2021) 153001},
  [\href{https://arxiv.org/abs/2103.01183}{{\ttfamily 2103.01183}}].

\bibitem{Riess:2021jrx}
A.~G. Riess et~al., \emph{{A Comprehensive Measurement of the Local Value of
  the Hubble Constant with 1 km/s/Mpc Uncertainty from the Hubble Space
  Telescope and the SH0ES Team}},
  \href{https://arxiv.org/abs/2112.04510}{{\ttfamily 2112.04510}}.

\bibitem{Sandage:2006cv}
A.~Sandage, G.~A. Tammann, A.~Saha, B.~Reindl, F.~D. Macchetto and N.~Panagia,
  \emph{{The Hubble Constant: A Summary of the HST Program for the Luminosity
  Calibration of Type Ia Supernovae by Means of Cepheids}},
  \href{https://doi.org/10.1086/508853}{\emph{Astrophys. J.} {\bfseries 653}
  (2006) 843--860}, [\href{https://arxiv.org/abs/astro-ph/0603647}{{\ttfamily
  astro-ph/0603647}}].

\bibitem{Freedman:2020dne}
W.~L. Freedman, B.~F. Madore, T.~Hoyt, I.~S. Jang, R.~Beaton, M.~G. Lee et~al.,
  \emph{{Calibration of the Tip of the Red Giant Branch (TRGB)}},
  \href{https://arxiv.org/abs/2002.01550}{{\ttfamily 2002.01550}}.

\bibitem{Huang:2019yhh}
C.~D. Huang, A.~G. Riess, W.~Yuan, L.~M. Macri, N.~L. Zakamska, S.~Casertano
  et~al., \emph{{Hubble Space Telescope Observations of Mira Variables in the
  Type Ia Supernova Host NGC 1559: An Alternative Candle to Measure the Hubble
  Constant}},  \href{https://arxiv.org/abs/1908.10883}{{\ttfamily 1908.10883}}.

\bibitem{Ivanov:2020ril}
M.~M. Ivanov, E.~McDonough, J.~C. Hill, M.~Simonovi\'c, M.~W. Toomey,
  S.~Alexander et~al., \emph{{Constraining Early Dark Energy with Large-Scale
  Structure}}, \href{https://doi.org/10.1103/PhysRevD.102.103502}{\emph{Phys.
  Rev. D} {\bfseries 102} (2020) 103502},
  [\href{https://arxiv.org/abs/2006.11235}{{\ttfamily 2006.11235}}].

\bibitem{Schoneberg:2021qvd}
N.~Sch\"oneberg, G.~Franco~Abell\'an, A.~P\'erez~S\'anchez, S.~J. Witte,
  V.~Poulin and J.~Lesgourgues, \emph{{The $H_0$ Olympics: A fair ranking of
  proposed models}},  \href{https://arxiv.org/abs/2107.10291}{{\ttfamily
  2107.10291}}.

\bibitem{Karwal:2016vyq}
T.~Karwal and M.~Kamionkowski, \emph{{Dark energy at early times, the Hubble
  parameter, and the string axiverse}},
  \href{https://doi.org/10.1103/PhysRevD.94.103523}{\emph{Phys. Rev. D}
  {\bfseries 94} (2016) 103523},
  [\href{https://arxiv.org/abs/1608.01309}{{\ttfamily 1608.01309}}].

\bibitem{Poulin:2018cxd}
V.~Poulin, T.~L. Smith, T.~Karwal and M.~Kamionkowski, \emph{{Early Dark Energy
  Can Resolve The Hubble Tension}},
  \href{https://doi.org/10.1103/PhysRevLett.122.221301}{\emph{Phys. Rev. Lett.}
  {\bfseries 122} (2019) 221301},
  [\href{https://arxiv.org/abs/1811.04083}{{\ttfamily 1811.04083}}].

\bibitem{Agrawal:2019lmo}
P.~Agrawal, F.-Y. Cyr-Racine, D.~Pinner and L.~Randall, \emph{{Rock 'n' Roll
  Solutions to the Hubble Tension}},
  \href{https://arxiv.org/abs/1904.01016}{{\ttfamily 1904.01016}}.

\bibitem{Alexander:2019rsc}
S.~Alexander and E.~McDonough, \emph{{Axion-Dilaton Destabilization and the
  Hubble Tension}},
  \href{https://doi.org/10.1016/j.physletb.2019.134830}{\emph{Phys. Lett. B}
  {\bfseries 797} (2019) 134830},
  [\href{https://arxiv.org/abs/1904.08912}{{\ttfamily 1904.08912}}].

\bibitem{Lin:2019qug}
M.-X. Lin, G.~Benevento, W.~Hu and M.~Raveri, \emph{{Acoustic Dark Energy:
  Potential Conversion of the Hubble Tension}},
  \href{https://doi.org/10.1103/PhysRevD.100.063542}{\emph{Phys. Rev. D}
  {\bfseries 100} (2019) 063542},
  [\href{https://arxiv.org/abs/1905.12618}{{\ttfamily 1905.12618}}].

\bibitem{Perez:2020cwa}
A.~Perez, D.~Sudarsky and E.~Wilson-Ewing, \emph{{Resolving the $H_0$ tension
  with diffusion}},
  \href{https://doi.org/10.1007/s10714-020-02781-0}{\emph{Gen. Rel. Grav.}
  {\bfseries 53} (2021) 7}, [\href{https://arxiv.org/abs/2001.07536}{{\ttfamily
  2001.07536}}].

\bibitem{Niedermann:2019olb}
F.~Niedermann and M.~S. Sloth, \emph{{New early dark energy}},
  \href{https://doi.org/10.1103/PhysRevD.103.L041303}{\emph{Phys. Rev. D}
  {\bfseries 103} (2021) L041303},
  [\href{https://arxiv.org/abs/1910.10739}{{\ttfamily 1910.10739}}].

\bibitem{Sakstein:2019fmf}
J.~Sakstein and M.~Trodden, \emph{{Early Dark Energy from Massive Neutrinos as
  a Natural Resolution of the Hubble Tension}},
  \href{https://doi.org/10.1103/PhysRevLett.124.161301}{\emph{Phys. Rev. Lett.}
  {\bfseries 124} (2020) 161301},
  [\href{https://arxiv.org/abs/1911.11760}{{\ttfamily 1911.11760}}].

\bibitem{Ye:2020btb}
G.~Ye and Y.-S. Piao, \emph{{Is the Hubble tension a hint of AdS phase around
  recombination?}},
  \href{https://doi.org/10.1103/PhysRevD.101.083507}{\emph{Phys. Rev. D}
  {\bfseries 101} (2020) 083507},
  [\href{https://arxiv.org/abs/2001.02451}{{\ttfamily 2001.02451}}].

\bibitem{Freese:2021rjq}
K.~Freese and M.~W. Winkler, \emph{{Chain early dark energy: A Proposal for
  solving the Hubble tension and explaining today\textquoteright{}s dark
  energy}}, \href{https://doi.org/10.1103/PhysRevD.104.083533}{\emph{Phys. Rev.
  D} {\bfseries 104} (2021) 083533},
  [\href{https://arxiv.org/abs/2102.13655}{{\ttfamily 2102.13655}}].

\bibitem{Braglia:2020bym}
M.~Braglia, W.~T. Emond, F.~Finelli, A.~E. Gumrukcuoglu and K.~Koyama,
  \emph{{Unified framework for early dark energy from $\alpha$-attractors}},
  \href{https://doi.org/10.1103/PhysRevD.102.083513}{\emph{Phys. Rev. D}
  {\bfseries 102} (2020) 083513},
  [\href{https://arxiv.org/abs/2005.14053}{{\ttfamily 2005.14053}}].

\bibitem{Gogoi:2020qif}
A.~Gogoi, R.~K. Sharma, P.~Chanda and S.~Das, \emph{{Early Mass-varying
  Neutrino Dark Energy: Nugget Formation and Hubble Anomaly}},
  \href{https://doi.org/10.3847/1538-4357/abfe5b}{\emph{Astrophys. J.}
  {\bfseries 915} (2021) 132},
  [\href{https://arxiv.org/abs/2005.11889}{{\ttfamily 2005.11889}}].

\bibitem{Turner:1983he}
M.~S. Turner, \emph{{Coherent Scalar Field Oscillations in an Expanding
  Universe}}, \href{https://doi.org/10.1103/PhysRevD.28.1243}{\emph{Phys. Rev.
  D} {\bfseries 28} (1983) 1243}.

\bibitem{Poulin:2018dzj}
V.~Poulin, T.~L. Smith, D.~Grin, T.~Karwal and M.~Kamionkowski,
  \emph{{Cosmological implications of ultralight axionlike fields}},
  \href{https://doi.org/10.1103/PhysRevD.98.083525}{\emph{Phys. Rev. D}
  {\bfseries 98} (2018) 083525},
  [\href{https://arxiv.org/abs/1806.10608}{{\ttfamily 1806.10608}}].

\bibitem{Smith:2019ihp}
T.~L. Smith, V.~Poulin and M.~A. Amin, \emph{{Oscillating scalar fields and the
  Hubble tension: a resolution with novel signatures}},
  \href{https://doi.org/10.1103/PhysRevD.101.063523}{\emph{Phys. Rev. D}
  {\bfseries 101} (2020) 063523},
  [\href{https://arxiv.org/abs/1908.06995}{{\ttfamily 1908.06995}}].

\bibitem{Hill:2020osr}
J.~C. Hill, E.~McDonough, M.~W. Toomey and S.~Alexander, \emph{{Early dark
  energy does not restore cosmological concordance}},
  \href{https://doi.org/10.1103/PhysRevD.102.043507}{\emph{Phys. Rev. D}
  {\bfseries 102} (2020) 043507},
  [\href{https://arxiv.org/abs/2003.07355}{{\ttfamily 2003.07355}}].

\bibitem{DAmico:2020ods}
G.~D'Amico, L.~Senatore, P.~Zhang and H.~Zheng, \emph{{The Hubble Tension in
  Light of the Full-Shape Analysis of Large-Scale Structure Data}},
  \href{https://doi.org/10.1088/1475-7516/2021/05/072}{\emph{JCAP} {\bfseries
  05} (2021) 072}, [\href{https://arxiv.org/abs/2006.12420}{{\ttfamily
  2006.12420}}].

\bibitem{Herold:2021ksg}
L.~Herold, E.~G.~M. Ferreira and E.~Komatsu, \emph{{New Constraint on Early
  Dark Energy from Planck and BOSS Data Using the Profile Likelihood}},
  \href{https://doi.org/10.3847/2041-8213/ac63a3}{\emph{Astrophys. J. Lett.}
  {\bfseries 929} (2022) L16},
  [\href{https://arxiv.org/abs/2112.12140}{{\ttfamily 2112.12140}}].

\bibitem{Smith:2020rxx}
T.~L. Smith, V.~Poulin, J.~L. Bernal, K.~K. Boddy, M.~Kamionkowski and
  R.~Murgia, \emph{{Early dark energy is not excluded by current large-scale
  structure data}},
  \href{https://doi.org/10.1103/PhysRevD.103.123542}{\emph{Phys. Rev. D}
  {\bfseries 103} (2021) 123542},
  [\href{https://arxiv.org/abs/2009.10740}{{\ttfamily 2009.10740}}].

\bibitem{Hill:2021yec}
J.~C. Hill et~al., \emph{{Atacama Cosmology Telescope: Constraints on
  prerecombination early dark energy}},
  \href{https://doi.org/10.1103/PhysRevD.105.123536}{\emph{Phys. Rev. D}
  {\bfseries 105} (2022) 123536},
  [\href{https://arxiv.org/abs/2109.04451}{{\ttfamily 2109.04451}}].

\bibitem{Vagnozzi:2021gjh}
S.~Vagnozzi, \emph{{Consistency tests of \ensuremath{\Lambda}CDM from the early
  integrated Sachs-Wolfe effect: Implications for early-time new physics and
  the Hubble tension}},
  \href{https://doi.org/10.1103/PhysRevD.104.063524}{\emph{Phys. Rev. D}
  {\bfseries 104} (2021) 063524},
  [\href{https://arxiv.org/abs/2105.10425}{{\ttfamily 2105.10425}}].

\bibitem{HSC:2018mrq}
{\scshape HSC} collaboration, C.~Hikage et~al., \emph{{Cosmology from cosmic
  shear power spectra with Subaru Hyper Suprime-Cam first-year data}},
  \href{https://doi.org/10.1093/pasj/psz010}{\emph{Publ. Astron. Soc. Jap.}
  {\bfseries 71} (2019) 43},
  [\href{https://arxiv.org/abs/1809.09148}{{\ttfamily 1809.09148}}].

\bibitem{KiDS:2020suj}
{\scshape KiDS} collaboration, M.~Asgari et~al., \emph{{KiDS-1000 Cosmology:
  Cosmic shear constraints and comparison between two point statistics}},
  \href{https://doi.org/10.1051/0004-6361/202039070}{\emph{Astron. Astrophys.}
  {\bfseries 645} (2021) A104},
  [\href{https://arxiv.org/abs/2007.15633}{{\ttfamily 2007.15633}}].

\bibitem{DES:2021vln}
{\scshape DES} collaboration, L.~F. Secco et~al., \emph{{Dark Energy Survey
  Year 3 results: Cosmology from cosmic shear and robustness to modeling
  uncertainty}}, \href{https://doi.org/10.1103/PhysRevD.105.023515}{\emph{Phys.
  Rev. D} {\bfseries 105} (2022) 023515},
  [\href{https://arxiv.org/abs/2105.13544}{{\ttfamily 2105.13544}}].

\bibitem{Nunes:2021ipq}
R.~C. Nunes and S.~Vagnozzi, \emph{{Arbitrating the S8 discrepancy with growth
  rate measurements from redshift-space distortions}},
  \href{https://doi.org/10.1093/mnras/stab1613}{\emph{Mon. Not. Roy. Astron.
  Soc.} {\bfseries 505} (2021) 5427--5437},
  [\href{https://arxiv.org/abs/2106.01208}{{\ttfamily 2106.01208}}].

\bibitem{Clark:2021hlo}
S.~J. Clark, K.~Vattis, J.~Fan and S.~M. Koushiappas, \emph{{The $H_0$ and
  $S_8$ tensions necessitate early and late time changes to $\Lambda$CDM}},
  \href{https://arxiv.org/abs/2110.09562}{{\ttfamily 2110.09562}}.

\bibitem{Allali:2021azp}
I.~J. Allali, M.~P. Hertzberg and F.~Rompineve, \emph{{Dark sector to restore
  cosmological concordance}},
  \href{https://doi.org/10.1103/PhysRevD.104.L081303}{\emph{Phys. Rev. D}
  {\bfseries 104} (2021) L081303},
  [\href{https://arxiv.org/abs/2104.12798}{{\ttfamily 2104.12798}}].

\bibitem{Ye:2021iwa}
G.~Ye, J.~Zhang and Y.-S. Piao, \emph{{Resolving both $H_0$ and $S_8$ tensions
  with AdS early dark energy and ultralight axion}},
  \href{https://arxiv.org/abs/2107.13391}{{\ttfamily 2107.13391}}.

\bibitem{Reeves:2022aoi}
A.~Reeves, L.~Herold, S.~Vagnozzi, B.~D. Sherwin and E.~G.~M. Ferreira,
  \emph{{Restoring cosmological concordance with early dark energy and massive
  neutrinos?}},  \href{https://arxiv.org/abs/2207.01501}{{\ttfamily
  2207.01501}}.

\bibitem{McDonough:2021pdg}
E.~McDonough, M.-X. Lin, J.~C. Hill, W.~Hu and S.~Zhou, \emph{{The Early Dark
  Sector, the Hubble Tension, and the Swampland}},
  \href{https://arxiv.org/abs/2112.09128}{{\ttfamily 2112.09128}}.

\bibitem{Berghaus:2019cls}
K.~V. Berghaus and T.~Karwal, \emph{{Thermal Friction as a Solution to the
  Hubble Tension}},
  \href{https://doi.org/10.1103/PhysRevD.101.083537}{\emph{Phys. Rev. D}
  {\bfseries 101} (2020) 083537},
  [\href{https://arxiv.org/abs/1911.06281}{{\ttfamily 1911.06281}}].

\bibitem{Berghaus:2022cwf}
K.~V. Berghaus and T.~Karwal, \emph{{Thermal Friction as a Solution to the
  Hubble and Large-Scale Structure Tensions}},
  \href{https://arxiv.org/abs/2204.09133}{{\ttfamily 2204.09133}}.

\bibitem{Aloni:2021eaq}
D.~Aloni, A.~Berlin, M.~Joseph, M.~Schmaltz and N.~Weiner, \emph{{A Step in
  Understanding the Hubble Tension}},
  \href{https://arxiv.org/abs/2111.00014}{{\ttfamily 2111.00014}}.

\bibitem{Joseph:2022jsf}
M.~Joseph, D.~Aloni, M.~Schmaltz, E.~N. Sivarajan and N.~Weiner, \emph{{A Step
  in Understanding the $S_8$ Tension}},
  \href{https://arxiv.org/abs/2207.03500}{{\ttfamily 2207.03500}}.

\bibitem{2011arXiv1104.2932L}
J.~{Lesgourgues}, \emph{{The Cosmic Linear Anisotropy Solving System (CLASS) I:
  Overview}}, {\emph{arXiv e-prints} (Apr., 2011) arXiv:1104.2932},
  [\href{https://arxiv.org/abs/1104.2932}{{\ttfamily 1104.2932}}].

\bibitem{2011JCAP...07..034B}
D.~{Blas}, J.~{Lesgourgues} and T.~{Tram}, \emph{{The Cosmic Linear Anisotropy
  Solving System (CLASS). Part II: Approximation schemes}},
  \href{https://doi.org/10.1088/1475-7516/2011/07/034}{\emph{JCAP} {\bfseries
  2011} (July, 2011) 034}, [\href{https://arxiv.org/abs/1104.2933}{{\ttfamily
  1104.2933}}].

\bibitem{Marsh:2010wq}
D.~J.~E. Marsh and P.~G. Ferreira, \emph{{Ultra-Light Scalar Fields and the
  Growth of Structure in the Universe}},
  \href{https://doi.org/10.1103/PhysRevD.82.103528}{\emph{Phys. Rev. D}
  {\bfseries 82} (2010) 103528},
  [\href{https://arxiv.org/abs/1009.3501}{{\ttfamily 1009.3501}}].

\bibitem{Amendola:2005ad}
L.~Amendola and R.~Barbieri, \emph{{Dark matter from an ultra-light
  pseudo-Goldsone-boson}},
  \href{https://doi.org/10.1016/j.physletb.2006.08.069}{\emph{Phys. Lett. B}
  {\bfseries 642} (2006) 192--196},
  [\href{https://arxiv.org/abs/hep-ph/0509257}{{\ttfamily hep-ph/0509257}}].

\bibitem{hu}
W.~Hu, \emph{Structure formation with generalized dark matter},
  \href{https://doi.org/10.1086/306274}{\emph{The Astrophysical Journal}
  {\bfseries 506} (Oct, 1998) 485–494}.

\bibitem{Hwang:2009js}
J.-c. Hwang and H.~Noh, \emph{{Axion as a Cold Dark Matter candidate}},
  \href{https://doi.org/10.1016/j.physletb.2009.08.031}{\emph{Phys. Lett. B}
  {\bfseries 680} (2009) 1--3},
  [\href{https://arxiv.org/abs/0902.4738}{{\ttfamily 0902.4738}}].

\bibitem{1974A&A....37..225M}
P.~{Meszaros}, \emph{{The behaviour of point masses in an expanding
  cosmological substratum.}}, {\emph{Astron. \& Astrophys.} {\bfseries 37}
  (Dec., 1974) 225--228}.

\bibitem{Hlozek:2014lca}
R.~Hlozek, D.~Grin, D.~J.~E. Marsh and P.~G. Ferreira, \emph{{A search for
  ultralight axions using precision cosmological data}},
  \href{https://doi.org/10.1103/PhysRevD.91.103512}{\emph{Phys. Rev. D}
  {\bfseries 91} (2015) 103512},
  [\href{https://arxiv.org/abs/1410.2896}{{\ttfamily 1410.2896}}].

\bibitem{Ooguri:2006in}
H.~Ooguri and C.~Vafa, \emph{{On the Geometry of the String Landscape and the
  Swampland}},
  \href{https://doi.org/10.1016/j.nuclphysb.2006.10.033}{\emph{Nucl. Phys. B}
  {\bfseries 766} (2007) 21--33},
  [\href{https://arxiv.org/abs/hep-th/0605264}{{\ttfamily hep-th/0605264}}].

\bibitem{Arkani-Hamed:2006emk}
N.~Arkani-Hamed, L.~Motl, A.~Nicolis and C.~Vafa, \emph{{The String landscape,
  black holes and gravity as the weakest force}},
  \href{https://doi.org/10.1088/1126-6708/2007/06/060}{\emph{JHEP} {\bfseries
  06} (2007) 060}, [\href{https://arxiv.org/abs/hep-th/0601001}{{\ttfamily
  hep-th/0601001}}].

\bibitem{DES:2021wwk}
{\scshape DES} collaboration, T.~M.~C. Abbott et~al., \emph{{Dark Energy Survey
  Year 3 results: Cosmological constraints from galaxy clustering and weak
  lensing}}, \href{https://doi.org/10.1103/PhysRevD.105.023520}{\emph{Phys.
  Rev. D} {\bfseries 105} (2022) 023520},
  [\href{https://arxiv.org/abs/2105.13549}{{\ttfamily 2105.13549}}].

\bibitem{Flitter:2022pzf}
J.~Flitter and E.~D. Kovetz, \emph{{Closing the window on fuzzy dark matter
  with the 21cm signal}},  \href{https://arxiv.org/abs/2207.05083}{{\ttfamily
  2207.05083}}.

\bibitem{Allali:2022yvx}
I.~J. Allali, M.~P. Hertzberg and Y.~Lyu, \emph{{Altered Axion Abundance from a
  Dynamical Peccei-Quinn Scale}},
  \href{https://doi.org/10.1103/PhysRevD.105.123517}{\emph{Phys. Rev. D}
  {\bfseries 105} (2022) 123517},
  [\href{https://arxiv.org/abs/2203.15817}{{\ttfamily 2203.15817}}].

\bibitem{Cicoli:2021gss}
M.~Cicoli, V.~Guidetti, N.~Righi and A.~Westphal, \emph{{Fuzzy Dark Matter
  candidates from string theory}},
  \href{https://doi.org/10.1007/JHEP05(2022)107}{\emph{JHEP} {\bfseries 05}
  (2022) 107}, [\href{https://arxiv.org/abs/2110.02964}{{\ttfamily
  2110.02964}}].

\bibitem{Bernardo:2022ztc}
H.~Bernardo, R.~Brandenberger and J.~Fr\"ohlich, \emph{{Towards a Dark Sector
  Model from String Theory}},
  \href{https://arxiv.org/abs/2201.04668}{{\ttfamily 2201.04668}}.

\bibitem{Burgess:2021obw}
C.~P. Burgess, D.~Dineen and F.~Quevedo, \emph{{Yoga Dark Energy: natural
  relaxation and other dark implications of a supersymmetric gravity sector}},
  \href{https://doi.org/10.1088/1475-7516/2022/03/064}{\emph{JCAP} {\bfseries
  03} (2022) 064}, [\href{https://arxiv.org/abs/2111.07286}{{\ttfamily
  2111.07286}}].

\bibitem{Grana:2005jc}
M.~Grana, \emph{{Flux compactifications in string theory: A Comprehensive
  review}}, \href{https://doi.org/10.1016/j.physrep.2005.10.008}{\emph{Phys.
  Rept.} {\bfseries 423} (2006) 91--158},
  [\href{https://arxiv.org/abs/hep-th/0509003}{{\ttfamily hep-th/0509003}}].

\bibitem{Blumenhagen:2006ci}
R.~Blumenhagen, B.~Kors, D.~Lust and S.~Stieberger, \emph{{Four-dimensional
  String Compactifications with D-Branes, Orientifolds and Fluxes}},
  \href{https://doi.org/10.1016/j.physrep.2007.04.003}{\emph{Phys. Rept.}
  {\bfseries 445} (2007) 1--193},
  [\href{https://arxiv.org/abs/hep-th/0610327}{{\ttfamily hep-th/0610327}}].

\bibitem{wess2020supersymmetry}
J.~Wess and J.~Bagger, \emph{Supersymmetry and Supergravity: Revised Edition},
  vol.~103.
\newblock Princeton university press, 2020.

\bibitem{freedman2012supergravity}
D.~Z. Freedman and A.~Van~Proeyen, \emph{Supergravity}.
\newblock Cambridge university press, 2012.

\bibitem{Lahanas:1986uc}
A.~B. Lahanas and D.~V. Nanopoulos, \emph{{The Road to No Scale Supergravity}},
  \href{https://doi.org/10.1016/0370-1573(87)90034-2}{\emph{Phys. Rept.}
  {\bfseries 145} (1987) 1}.

\bibitem{Cicoli:2013rwa}
M.~Cicoli, S.~de~Alwis and A.~Westphal, \emph{{Heterotic Moduli
  Stabilisation}}, \href{https://doi.org/10.1007/JHEP10(2013)199}{\emph{JHEP}
  {\bfseries 10} (2013) 199},
  [\href{https://arxiv.org/abs/1304.1809}{{\ttfamily 1304.1809}}].

\bibitem{Svrcek:2006yi}
P.~Svrcek and E.~Witten, \emph{{Axions In String Theory}},
  \href{https://doi.org/10.1088/1126-6708/2006/06/051}{\emph{JHEP} {\bfseries
  06} (2006) 051}, [\href{https://arxiv.org/abs/hep-th/0605206}{{\ttfamily
  hep-th/0605206}}].

\bibitem{Polchinski:1998rr}
J.~Polchinski, \emph{{String theory. Vol. 2: Superstring theory and beyond}}.
\newblock Cambridge Monographs on Mathematical Physics. Cambridge University
  Press, 12, 2007,
  \href{https://doi.org/10.1017/CBO9780511618123}{10.1017/CBO9780511618123}.

\bibitem{Arvanitaki:2009fg}
A.~Arvanitaki, S.~Dimopoulos, S.~Dubovsky, N.~Kaloper and J.~March-Russell,
  \emph{{String Axiverse}},
  \href{https://doi.org/10.1103/PhysRevD.81.123530}{\emph{Phys. Rev. D}
  {\bfseries 81} (2010) 123530},
  [\href{https://arxiv.org/abs/0905.4720}{{\ttfamily 0905.4720}}].

\bibitem{Cicoli:2012sz}
M.~Cicoli, M.~Goodsell and A.~Ringwald, \emph{{The type IIB string axiverse and
  its low-energy phenomenology}},
  \href{https://doi.org/10.1007/JHEP10(2012)146}{\emph{JHEP} {\bfseries 10}
  (2012) 146}, [\href{https://arxiv.org/abs/1206.0819}{{\ttfamily 1206.0819}}].

\bibitem{Dasgupta:2008hb}
K.~Dasgupta, H.~Firouzjahi and R.~Gwyn, \emph{{On The Warped Heterotic Axion}},
  \href{https://doi.org/10.1088/1126-6708/2008/06/056}{\emph{JHEP} {\bfseries
  06} (2008) 056}, [\href{https://arxiv.org/abs/0803.3828}{{\ttfamily
  0803.3828}}].

\bibitem{Franco:2014hsa}
S.~Franco, D.~Galloni, A.~Retolaza and A.~Uranga, \emph{{On axion monodromy
  inflation in warped throats}},
  \href{https://doi.org/10.1007/JHEP02(2015)086}{\emph{JHEP} {\bfseries 02}
  (2015) 086}, [\href{https://arxiv.org/abs/1405.7044}{{\ttfamily 1405.7044}}].

\bibitem{McDonough:2018xzh}
E.~McDonough and S.~Alexander, \emph{{Observable Chiral Gravitational Waves
  from Inflation in String Theory}},
  \href{https://doi.org/10.1088/1475-7516/2018/11/030}{\emph{JCAP} {\bfseries
  11} (2018) 030}, [\href{https://arxiv.org/abs/1806.05684}{{\ttfamily
  1806.05684}}].

\bibitem{Hebecker:2018yxs}
A.~Hebecker, S.~Leonhardt, J.~Moritz and A.~Westphal, \emph{{Thraxions:
  Ultralight Throat Axions}},
  \href{https://doi.org/10.1007/JHEP04(2019)158}{\emph{JHEP} {\bfseries 04}
  (2019) 158}, [\href{https://arxiv.org/abs/1812.03999}{{\ttfamily
  1812.03999}}].

\bibitem{Ma:1995ey}
C.-P. Ma and E.~Bertschinger, \emph{{Cosmological perturbation theory in the
  synchronous and conformal Newtonian gauges}},
  \href{https://doi.org/10.1086/176550}{\emph{Astrophys. J.} {\bfseries 455}
  (1995) 7--25}, [\href{https://arxiv.org/abs/astro-ph/9506072}{{\ttfamily
  astro-ph/9506072}}].

\end{thebibliography}\endgroup

\end{document}